\documentclass[aps,manuscript]{revtex4}
\usepackage[T1]{fontenc}
\usepackage[latin9]{inputenc}
\usepackage{float}
\usepackage{amstext}
\usepackage{amssymb}
\usepackage{graphicx}
\usepackage{esint}

\makeatletter
\@ifundefined{textcolor}{}
{%
 \definecolor{BLACK}{gray}{0}
 \definecolor{WHITE}{gray}{1}
 \definecolor{RED}{rgb}{1,0,0}
 \definecolor{GREEN}{rgb}{0,1,0}
 \definecolor{BLUE}{rgb}{0,0,1}
 \definecolor{CYAN}{cmyk}{1,0,0,0}
 \definecolor{MAGENTA}{cmyk}{0,1,0,0}
 \definecolor{YELLOW}{cmyk}{0,0,1,0}
}

\@ifundefined{showcaptionsetup}{}{%
 \PassOptionsToPackage{caption=false}{subfig}}
\usepackage{subfig}
\makeatother

\usepackage[english]{babel}
\begin{document}

\title{Effects of interactions on the dynamics of driven cold atoms}

\author{Alexandra Bakman and Shmuel Fishman}
\begin{abstract}
The quantum fidelity was introduced by Peres to study some fingerprints
of classically chaotic behavior in the quantum dynamics of the corresponding
systems. In the present paper the signatures of classical dynamics
near elliptic points and of interactions between particles are characterized
for kicked systems. In particular, the period of the fidelity resulting
of the interactions is found using analytical and numerical calculations.
A mechanism leading to the oscillations with the intermediate period
is proposed. It is of a semiclassical origin and results of the interplay
between the oscillations of the width of the wave packets and the
rotation of their center around the elliptic fixed point. 
\end{abstract}
\maketitle

\section{Introduction}

Effects of inter-particle interactions on the dynamics of driven systems
were the subject of several recent works \cite{Quantum_accelerator,Transition_to_instability,Wave_packet_dynamics,Mark_paper}.
In the present paper these studies will be extended to the exploration
of the effects of interactions on the quantum fidelity.

The concept of quantum fidelity was introduced by Peres \cite{Peres_fidelity}
as a fingerprint of classical chaos in quantum dynamics. It has subsequently
been extensively utilized in theoretical \cite{Jalabert_Pastawski_Y19,Decay_Loschmidt_Y20,Sensitivity_Chaotic_Systems_Y34,Saturation_of_Fidelity_Y35}
and experimental studies \cite{Saturation_of_Fidelity_Y35,Decay_of_QM_correlations_Y38,Hyperfine_spectroscopy_Y37,Revivals_of_coherence_Y36}
(for a review see \cite{Review_Y21}). In absence of interactions
the quantum fidelity, in a mixed system (in some parts of phase space
the dynamics is chaotic and in other parts regular) was studied \cite{Yevgeny_paper}.
In particular, it was found that the fidelity exhibits oscillations
in time, and their periods are found to be related to the periods
of the motion in regular parts of phase space \cite{Yevgeny_paper}. 

In the present work we will study the effects of the inter-particle
interactions on the periods of the fidelity. The fidelity is defined
by 

\begin{equation}
F\left(t\right)=|\left\langle \psi_{1}|\psi_{2}\right\rangle |^{2}\label{eq:Fidelity bra ket form}
\end{equation}

\noindent where 

\begin{equation}
\left|\psi_{1}\left(t\right)\right\rangle =e^{i\frac{H_{1}t}{\hbar}}\left|\phi_{0}\right\rangle 
\end{equation}

\noindent and

\begin{equation}
\left|\psi_{2}\left(t\right)\right\rangle =e^{i\frac{H_{2}t}{\hbar}}\left|\phi_{0}\right\rangle 
\end{equation}

\noindent are propagated by the Hamiltonians $H_{1}$ and $H_{2}$,
that are of the same form but with different values of the parameters
and $\left|\phi_{0}\right\rangle $ is the initial state.

We note that the fidelity $F\left(t\right)$ is related to the integral
over Wigner functions,

\begin{equation}
F\left(t\right)=\intop_{-\infty}^{\infty}\intop_{-\infty}^{\infty}dxdpW_{1}\left(x,p\right)W_{2}\left(x,p\right)\label{eq:Fidelity using Wigner functions}
\end{equation}

\noindent where $W_{1}$ and $W_{2}$ are the Wigner functions of
$\left|\psi_{1}\right\rangle $ and $\left|\psi_{2}\right\rangle $
,respectively. 

The general form of the Wigner function is

\begin{equation}
W\left(x,p\right)=\frac{1}{\pi\cdot\hbar}\intop_{-\infty}^{\infty}d\xi\cdot\psi^{*}\left(x+\xi\right)\psi\left(x-\xi\right)e^{\frac{2ip\xi}{\hbar}}.\label{eq:Wigner function integral form}
\end{equation}

\noindent Without interactions, the specific system we will study
is defined by the Hamiltonian \cite{Yevgeny_paper}

\begin{equation}
H=\frac{p^{2}}{2}-Ke^{-\frac{x^{2}}{2}}\sum_{n=-\infty}^{\infty}\delta\left(t-n\right)\label{eq:Rescaled Hamiltonian without interactions}
\end{equation}

\noindent where

\begin{equation}
p=-i\tau\partial_{x}
\end{equation}

\noindent and

\begin{equation}
\tau=\frac{\hbar T}{m\Delta^{2}}\label{eq:tau}
\end{equation}

\noindent is the rescaled $\hbar$ ,satisfying 

\begin{equation}
\left[x,p\right]=i\tau\label{eq:commutation relation}
\end{equation}

\noindent The Hamiltonian is in dimensionless units. In physical units
$T$ is the time between the kicks, $\triangle$ is the width of the
pulses of the kicking potential, while $m$ is the mass of the particles. 

\noindent The one step evolution operator is 

\begin{equation}
U=e^{-i\frac{p^{2}}{2\tau}}\exp\left(i\frac{K}{\tau}e^{-\frac{x^{2}}{2}}\right).\label{eq:One kick propagator without interactions}
\end{equation}

\noindent The corresponding classical map is 

\begin{equation}
p_{n+1}=p_{n}-Kx_{n}e^{-\frac{x_{n}^{2}}{2}}\label{eq:Classical map for momentum}
\end{equation}

\begin{equation}
x_{n+1}=x_{n}+p_{n+1}.\label{eq:Classical map for position}
\end{equation}

\begin{figure}
\centering{}\includegraphics[scale=0.7]{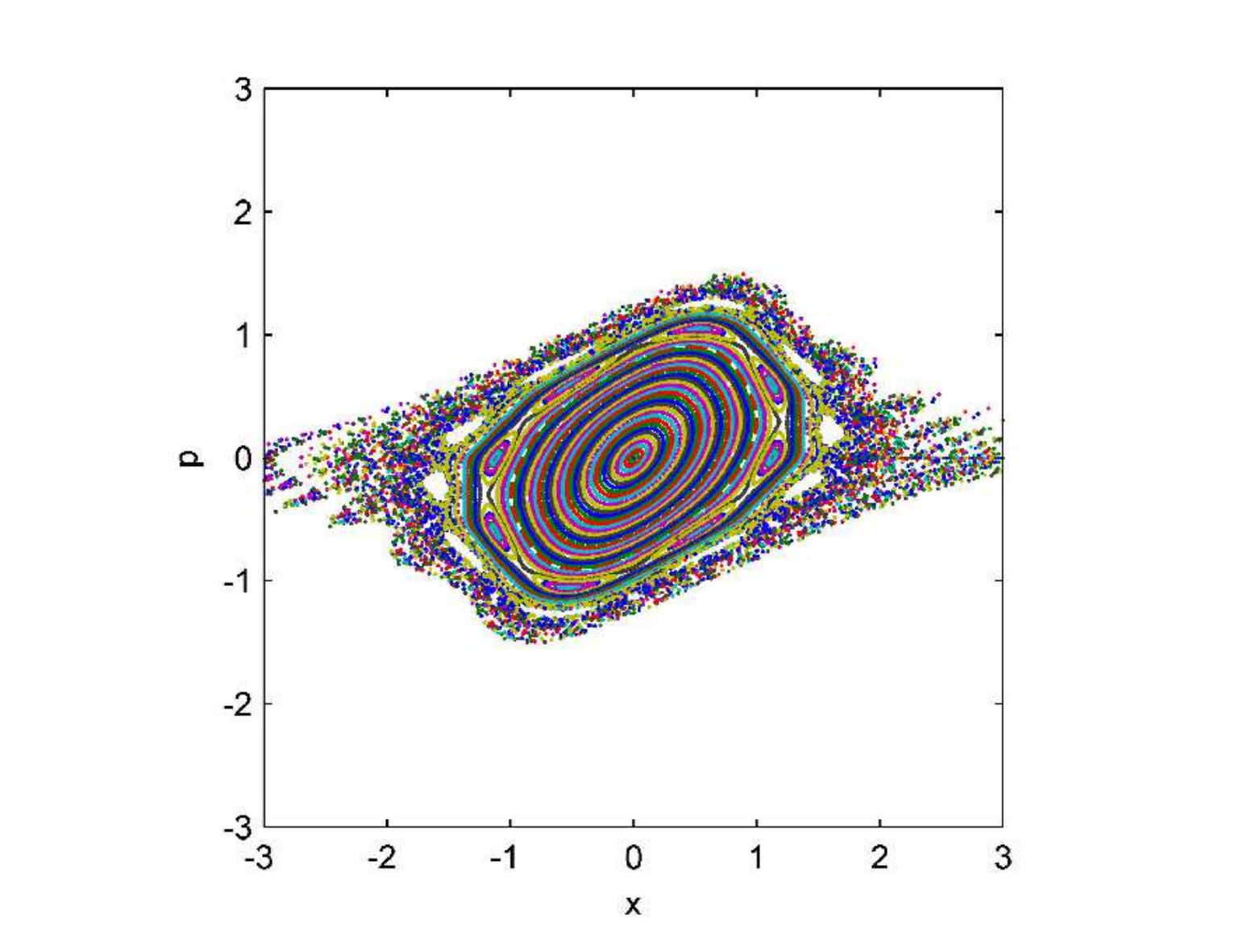}\caption{\label{fig:Phase space for K=00003D1}(Color online) The phase portrait
for $K=1$. Colors distinguish different orbits.}
\end{figure}

\noindent Its phase portrait is shown in Fig. \ref{fig:Phase space for K=00003D1}. 

In previous explorations \cite{Mark_Update} the interaction term
was introduced only between the kicks and the $\frac{p^{2}}{2}$ term
was replaced by

\begin{equation}
H_{I}=\frac{p^{2}}{2}+\beta\left|\psi\left(x\right)\right|^{2}\label{eq:Hamiltonian with interaction replacement}
\end{equation}

\noindent where $\beta$ is the strength of the interactions. Therefore,
between the kicks the dynamics are modeled by the nonlinear Schrödinger
equation (NLSE), known also as the Gross Pitaevskii equation (GPE),

\begin{equation}
i\tau\frac{\partial\psi}{\partial t}=H_{I}\psi.\label{eq:Schrodinger equations with interactions between kicks}
\end{equation}

In the expression for the evolution operator $e^{-\frac{p^{2}}{2\tau}}$
should be replaced by another evolution operator. In the calculation
of the fidelity \cite{Mark_Update}, the frequencies that were found
in the absence of interactions were observed. In addition, a different
new frequency was found. Unlike the other frequencies, this frequency
is not related in any simple way to the frequencies of the underlying
classical system. It was found to depend on the strength inter-particle
interactions and can be considered as a signature of the interactions
in the fidelity. 

The main problem with introducing the interaction term between kicks
\cite{Mark_Update} is that it requires the numerical solution of
the NLSE in each interval between kicks. This process is highly time
consuming since it requires the solution of a differential equation
between the kicks, and it is impossible to propagate the system for
very long times. 

For this reason, in the current work we study a model that is defined
by the evolution operator 

\begin{equation}
U=e^{-i\frac{p^{2}}{2\tau}}\exp\left(\frac{i}{\tau}\text{\ensuremath{\left(Ke^{-\frac{x^{2}}{2}}+\beta\left|\psi\left(x\right)\right|^{2}\right)}}\right)\label{eq:One kick propagator}
\end{equation}

\noindent where the interactions are introduced at the kicks. The
Hamiltonian of this model is

\begin{equation}
H=\frac{p^{2}}{2}-Ke^{-\frac{x^{2}}{2}}\sum_{n=-\infty}^{\infty}\delta\left(t-n\right)+\beta\left|\psi\left(x\right)\right|^{2}\sum_{n=-\infty}^{\infty}\delta\left(t-n\right).\label{eq:Hamiltonian with interactions}
\end{equation}
This model is related to one studied by Shepelyansky \cite{Shepelyansky_93}. 

It should be clarified that the purpose of this study is purely theoretical,
with the aim to shed light on the fingerprints of interparticle interactions
in the fidelity. We will focus our studies on the fidelity for wavepackets
started near the central elliptic fixed point. 

In Section \ref{sec:A model for motion near the central elliptic point}
we will introduce a harmonic oscillator model describing the motion
near the fixed point and discuss the modifications required. In Section
\ref{sec:Fidelity for weak interactions} we will introduce an approximate
theory for the fidelity oscillations and in Sections \ref{sec:Fidelity-oscillations}
and \ref{sec:The-origins-of_intermediate_period} we will confront
it with numerical results. The results are summarized and discussed
in Section \ref{sec:summary-and-conclusions}.

\section{A model for the motion near the central elliptic point\label{sec:A model for motion near the central elliptic point}}

Near the fixed point $\left(x,p\right)=\left(0,0\right)$, the dynamics
are approximately determined by the tangent map of the fixed point.
For this purpose we linearize the classical map (\ref{eq:Classical map for momentum}),
(\ref{eq:Classical map for position}) around the fixed point $x=p=0$.
This gives the equation for the deviations from this point

\begin{equation}
\left(\begin{array}{c}
\delta x_{n+1}\\
\delta p_{n+1}
\end{array}\right)=\left(\begin{array}{cc}
\left(1-K\right) & 1\\
-K & 1
\end{array}\right)\left(\begin{array}{c}
\delta x_{n}\\
\delta p_{n}
\end{array}\right).\label{eq:Tangent_map}
\end{equation}

\noindent The eigenvalues of this map are

\begin{equation}
\alpha_{\pm}=\frac{\left(2-K\right)}{2}\pm\frac{\sqrt{K\left(4-K\right)}}{2}\equiv e^{\pm i\omega}\label{eq:eigenvalues of tangent map}
\end{equation}

\noindent with

\begin{equation}
\omega=\arctan\left(\frac{\sqrt{K\left(4-K\right)}}{2-K}\right),\label{eq:Frequency of rotation around fixed point}
\end{equation}

\noindent which is the angular velocity of the points around the origin.
In the vicinity of the fixed point, the system behaves like a harmonic
oscillator with a frequency $\omega$. Classically, the motion of
the trajectories, starting near the elliptic fixed point, $x=p=0$,
stays there because the region is bounded by KAM curves that surround
this point. We consider here two Hamiltonians $H_{1}$ and $H_{2}$
that differ only by the values of the stochasticity parameter $K$,
taking the values $K_{1}=1$ and $K_{2}=1.01$. 

\noindent For $K=K_{1}=1$, one finds

\begin{equation}
\omega_{1}=1.047\label{eq:Explicit omega_1}
\end{equation}

\noindent and for $K=K_{2}=1.01$, 

\begin{equation}
\omega_{2}=1.053.\label{eq:Explicit omega_2}
\end{equation}

The parameters were chosen so that the map has a pronounced central
island as shown in Fig. \ref{fig:Phase space for K=00003D1}. The
qualititative behavior should be similar for all $0<K<4$ (see \cite{Yevgeny_paper}). 

The periods of the regular trajectories deviate from the ones found
at the elliptic point by (\ref{eq:Frequency of rotation around fixed point}).
The deviation increases with the deviation from the elliptic point.
This is similar to the situation when a small anharmonicity is added
to the harmonic potential. Therefore, for wave packets initiated not
exactly at the elliptic point, one has to add an anharmonic term to
model the dynamics. The result is that the different parts of the
packet are exposed to different frequencies. Therefore, an initially
prepared Gaussian wave packet spreads in phase space. This was indeed
verified for Gaussian wave packets in a harmonic well with a small
anharmonic correction \cite{Echoes}. As the wave packet propagates,
revivals are found for a very long time. Fortunately, in presence
of interactions, localization of Gaussian wave packets is possible,
as was found for an anharmonic well with inter-particle interactions
modeled by the Gross Pitaevskii equation (GPE) \cite{Mark_paper}
(see also \cite{Wave_packet_dynamics}). Interactions and nonlinearity
may balance each other to preserve the Gaussian wave packet \cite{Mark_paper}. 

In the following section, the dynamics of particles in a harmonic
well with a small anharmonic perturbation will be studied analytically,
for weak inter-particle interactions. Following the discussion in
the present section, it will be assumed that in the vicinity of this
elliptic point the motion can be described by a Gaussian wave packet.

\section{Fidelity for weak interactions\label{sec:Fidelity for weak interactions}}

In this section an estimate for the oscillations of the fidelity for
a wavepacket that is initially a coherent state of a harmonic oscillator
defined by the Hamiltonian

\begin{equation}
H=\frac{p^{2}}{2m}+\frac{1}{2}m\omega^{2}x^{2}.\label{eq:Hamiltonian of an harmonic oscillator}
\end{equation}

\subsection{The Wigner function of a coherent state}

A coherent state for the harmonic oscillator defined by the Hamiltonian
(\ref{eq:Hamiltonian of an harmonic oscillator}) is \cite{coherent_comp}

\begin{equation}
\psi_{0}=\left(\frac{m\omega}{\pi\hbar}\right)^{\frac{1}{4}}\exp\left(\frac{i}{\hbar}p_{0}\left(t\right)-\frac{m\omega}{2\hbar}\left(x-x_{0}\left(t\right)\right)^{2}\right)e^{-\frac{i}{2\hbar}x_{0}\cdot p_{0}}e^{-i\frac{\omega t}{2}}\label{eq: Coherent state wave function}
\end{equation}

\noindent where $x_{0}\left(t\right)$and $p_{0}\left(t\right)$ denote
the classical trajectory in phase space. The state (\ref{eq: Coherent state wave function})
is an eigenstate of the annihilation operator and satisfies the time
dependent Schrödinger's equation

\begin{equation}
i\hbar\frac{\partial\psi_{0}}{\partial t}=-\frac{\hbar^{2}}{2m}\frac{\partial^{2}\psi_{0}}{\partial x^{2}}+\frac{1}{2}m\omega^{2}x^{2}\psi_{0}\label{eq:Schrodinger's equation harmonic oscillator}
\end{equation}
The Wigner function of this coherent state is found from the definition
(\ref{eq:Wigner function integral form}):

\begin{equation}
W_{0}\left(x,p\right)=\frac{1}{\pi\cdot\hbar}e^{-\frac{m\omega}{\hbar}\left(x-x_{0}\right)^{2}}e^{-\frac{\left(p-p_{0}\right)^{2}}{m\omega\hbar}}\label{eq:Wigner function coherent state}
\end{equation}

\subsection{The fidelity for coherent states in absence of interactions}

Let $\omega_{1}$ and $\omega_{2}$ be the frequencies of two harmonic
oscillators, whose potentials differ slightly. The Wigner functions
for these wavefunctions are $\left(i=1,2\right)$

\begin{equation}
W_{i}\left(x,p\right)=\frac{1}{2\pi\sigma_{x_{i}}\sigma_{p_{i}}}\exp\left(-\frac{1}{2}\left(\frac{\left(x-x_{i}\left(t\right)\right)^{2}}{\sigma_{x_{i}}^{2}}+\frac{\left(p-p_{i}\left(t\right)\right)^{2}}{\sigma_{p_{i}}^{2}}\right)\right)\label{eq:Wigner function, harm. osc. general form}
\end{equation}

\noindent where

\begin{equation}
\sigma_{x_{i}}^{2}=\frac{\hbar}{2m\omega_{i}}\label{eq:width of Wigner function, x}
\end{equation}

\noindent and

\begin{equation}
\sigma_{p_{i}}^{2}=\frac{m\omega_{i}\hbar}{2}\label{eq:width of Wigner function, p}
\end{equation}

\noindent The fidelity in absence of interactions is calculated using
(\ref{eq:Fidelity using Wigner functions}) and is given by

\begin{equation}
F=Ce^{-\frac{1}{2}\left(s_{x}+s_{p}\right)}\label{eq:Fidelity, general form - Wigner functions}
\end{equation}

\noindent where the parameters are given by 

\begin{equation}
C=\frac{2}{\pi\hbar^{2}}\sqrt{\frac{\sigma_{x_{1}}^{2}\sigma_{x_{2}}^{2}\sigma_{p_{1}}^{2}\sigma_{p_{2}}^{2}}{\left(\sigma_{x_{1}}^{2}+\sigma_{x_{2}}^{2}\right)\left(\sigma_{p_{1}}^{2}+\sigma_{p_{2}}^{2}\right)}}\label{eq:C for Wigner fidelity}
\end{equation}

\begin{equation}
s_{x}=\frac{\left(x_{1}\left(t\right)-x_{2}\left(t\right)\right)^{2}}{\sigma_{x_{1}}^{2}+\sigma_{x_{2}}^{2}}\label{eq:s_x}
\end{equation}

\begin{equation}
s_{p}=\frac{\left(p_{1}\left(t\right)-p_{2}\left(t\right)\right)^{2}}{\sigma_{p_{1}}^{2}+\sigma_{p_{2}}^{2}}\label{eq:s_p}
\end{equation}

\noindent The classical trajectories are given by 

\begin{equation}
\left(x_{1},p_{1}\right)=\rho\left(\cos\left(\omega_{1}t\right),-m\omega_{1}\sin\left(\omega_{1}t\right)\right)\label{eq:classical trajectory, harmonic oscillator}
\end{equation}

\noindent and

\begin{equation}
\left(x_{2},p_{2}\right)=\rho\left(\cos\left(\omega_{2}t\right),-m\omega_{2}\sin\left(\omega_{2}t\right)\right)\label{eq:classical trajectory, harmonic oscillator-2}
\end{equation}

\noindent where

\begin{equation}
\rho=x_{1}\left(0\right)=x_{2}\left(0\right)
\end{equation}

\noindent Therefore, (\ref{eq:s_x}) can be written in the form

\begin{equation}
s_{x}=\frac{\rho^{2}}{\sigma_{x_{1}}^{2}+\sigma_{x_{2}}^{2}}\left(1+\frac{1}{2}\left(\cos\left(2\omega_{1}t\right)+\cos\left(2\omega_{2}t\right)\right)-\cos\left(\delta\omega\cdot t\right)-\cos\left(\omega_{s}t\right)\right)\label{eq:s_x with frequencies}
\end{equation}

\noindent where 

\begin{equation}
\omega_{s}=\omega_{1}+\omega_{2}\label{eq:w_s}
\end{equation}

\noindent and

\begin{equation}
\delta\omega=\omega_{1}-\omega_{2}\label{eq:delta_w}
\end{equation}

\noindent Similarly, 

\begin{eqnarray}
s_{p} & = & \frac{\rho^{2}m^{2}}{\sigma_{p_{1}}^{2}+\sigma_{p_{2}}^{2}}\left(\frac{\omega_{1}^{2}+\omega_{2}^{2}}{2}-\frac{1}{2}\left(\omega_{1}^{2}\cos\left(2\omega_{1}t\right)+\omega_{2}^{2}\cos\left(2\omega_{2}t\right)\right)\right)+\label{eq:s_p with frequencies}\\
 &  & +\frac{\rho^{2}m^{2}}{\sigma_{p_{1}}^{2}+\sigma_{p_{2}}^{2}}\left(\omega_{1}\omega_{2}\cos\left(\delta\omega\cdot t\right)-\omega_{1}\omega_{2}\cos\left(\omega_{s}t\right)\right)\nonumber 
\end{eqnarray}

\noindent For the model (\ref{eq:Rescaled Hamiltonian without interactions})
we study here, for $K_{1}=1$ and $K_{2}=1.01$, we find from (\ref{eq:Explicit omega_1})
and (\ref{eq:Explicit omega_2}) that 

\begin{equation}
\delta\omega=0.0057747\label{eq:delta_omega numerical value}
\end{equation}

\subsection{The fidelity for coherent states with weak interactions\label{sub:Fidelity for coherent states with weak interactions}}

We assume that the main effect of interactions is on the width of
the wave packets. 

\noindent The width of the wave packet is defined as

\begin{equation}
\left\langle \Delta x\right\rangle ^{2}=\left\langle \left(x-\left\langle x\right\rangle \right)^{2}\right\rangle ,\label{eq:width of wave packet, general expression}
\end{equation}

\noindent where $\left\langle O\right\rangle =\int_{-\infty}^{\infty}dx\psi^{*}O\psi$. 

Since we assume that the interactions are weak, the resulting correction
is expected to be small. We assume that the variation is periodic,
with a period $\Omega_{i}$ close to $2\omega_{i}$, an assumption
that will be verified numerically. A motivation for such an assumption
is that the expression for the width (\ref{eq:width of wave packet, general expression})
involves only the combinations of frequencies $\omega_{1}\pm\omega_{2}$,
$2\omega_{1}$ and $2\omega_{2}$. Following this assumption, we replace
(\ref{eq:width of Wigner function, x}) and (\ref{eq:width of Wigner function, p})
by

\begin{equation}
\tilde{\sigma}_{x_{1}}^{2}=\sigma_{x}^{2}+\gamma_{x}\cos\left(\Omega_{1}t+\phi_{x}\right)\label{eq:width in x with oscillations}
\end{equation}

\begin{equation}
\widetilde{\sigma}_{x_{2}}^{2}=\sigma_{x}^{2}+\gamma_{x}\cos\left(\Omega_{2}t+\phi_{x}\right)\label{eq:width in x with oscillations 2}
\end{equation}

\noindent and

\begin{equation}
\tilde{\sigma}_{p_{1}}^{2}=\sigma_{p}^{2}+\gamma_{p}\cos\left(\Omega_{1}t+\phi_{p}\right)\label{eq:width in p with oscillations}
\end{equation}

\begin{equation}
\widetilde{\sigma}_{p_{2}}^{2}=\sigma_{p}^{2}+\gamma_{p}\cos\left(\Omega_{2}t+\phi_{p}\right),\label{eq:with in p with oscillations 2}
\end{equation}

\noindent resulting in

\begin{equation}
s_{x}=\frac{\rho^{2}}{2\sigma_{x}^{2}}\left(\cos\left(\omega_{1}t\right)-\cos\left(\omega_{2}t\right)\right)^{2}\left(1+\frac{\gamma_{x}}{2\sigma_{x}^{2}}\left(\cos\left(\Omega_{1}t+\phi_{x}\right)+\cos\left(\Omega_{2}+\phi_{x}\right)\right)\right)^{-1}.\label{eq:s_x with oscillations}
\end{equation}

\noindent Similarly for $s_{p}$,

\begin{equation}
s_{p}=\frac{\rho^{2}m^{2}}{2\sigma_{p}^{2}}\left(\omega_{1}\sin\left(\omega_{1}t\right)+\omega_{2}\sin\left(\omega_{2}t\right)\right)^{2}\left(1+\frac{\gamma_{p}}{2\sigma_{p}^{2}}\left(\cos\left(\Omega_{1}t+\phi_{p}\right)+\cos\left(\Omega_{2}t+\phi_{p}\right)\right)\right)^{-1}.\label{eq:s_p with oscillations}
\end{equation}

\noindent We assume that the corrections resulting of the interactions
are small, therefore even with the replacement $\sigma\rightarrow\tilde{\sigma}$
the states $\psi_{i}$ are within a good approximation similar to
coherent states. 

\noindent Assuming $\left|\frac{\gamma_{x}}{2\sigma_{x}^{2}}\right|\ll1$,
in the leading order in $\frac{\gamma_{x}}{2\sigma_{x}^{2}},$ one
can simplify the expression as it is done in Appendix A. 

Our crucial assumption is that the Wigner function is well approximated
by a Gaussian wave packet. In the presence of interactions it is possible
that such a wave packet is stable in spite of the effective anharmonicity
generated for kicked systems, defined by (\ref{eq:Rescaled Hamiltonian without interactions})
as well as by (\ref{eq:One kick propagator}) - (\ref{eq:Hamiltonian with interactions})
(see \cite{Mark_paper}). In our case, where the interactions are
weak, the frequency of the width oscillation satisfies

\begin{equation}
\Omega_{1}\sim\Omega_{2}\equiv\Omega,\label{eq:omega1 like omega2}
\end{equation}

\begin{equation}
\omega_{1}\sim\omega_{2}\equiv\omega\label{eq:omega1 like omega2 (small)}
\end{equation}

\noindent and

\begin{equation}
\Omega_{1,2}\sim\omega_{1,2}\gg\delta\omega.\label{eq:Omega >>delta_omega}
\end{equation}

\noindent We denote

\begin{equation}
\omega_{s}=\omega_{1}+\omega_{2}\sim2\omega.\label{eq:w_s like 2w}
\end{equation}

\noindent Using the approximation in (\ref{eq:w_s like 2w}), we denote

\begin{equation}
\Delta\omega=\omega_{s}-\Omega\simeq2\omega-\Omega.\label{eq:D_omega}
\end{equation}

\noindent we find (see Appendix A)

\begin{equation}
s_{x}+s_{p}=\sum_{i=1}^{8}A_{i}\label{eq:s_x+s_p=00003Dsum}
\end{equation}

\noindent where

\begin{equation}
A_{1}=\frac{\rho^{2}}{2\sigma_{x}^{2}}+\frac{\rho^{2}m^{2}\omega^{2}}{2\sigma_{p}^{2}},\label{eq:A1}
\end{equation}

\begin{equation}
A_{2}=\left(-\frac{\rho^{2}}{2\sigma_{x}^{2}}+\frac{\rho^{2}m^{2}\omega^{2}}{2\sigma_{p}^{2}}\right)\cos\left(\delta\omega\cdot t\right),\label{eq:A2}
\end{equation}

\begin{equation}
A_{3}=-\frac{\rho^{2}m^{2}\omega^{2}}{\sigma_{p}^{2}}\cos\left(\omega_{s}\cdot t\right),\label{eq:A3}
\end{equation}

\begin{equation}
A_{4}=\frac{\rho^{2}\gamma_{x}}{2\sigma_{x}^{4}}\cos\left(\Omega\cdot t+\phi_{x}\right)-\frac{\rho^{2}\gamma_{p}m^{2}\omega^{2}}{2\sigma_{p}^{4}}\cos\left(\Omega\cdot t+\phi_{p}\right),\label{eq:A4}
\end{equation}

\begin{equation}
A_{5}=\frac{\rho^{2}\gamma_{x}}{4\sigma_{x}^{4}}\cos\left(\left(\Omega-\delta\omega\right)t+\phi_{x}\right)-\frac{\rho^{2}m^{2}\gamma_{p}\omega^{2}}{4\sigma_{p}^{4}}\cos\left(\left(\Omega-\delta\omega\right)t+\phi_{p}\right),\label{eq:A5}
\end{equation}

\begin{equation}
A_{6}=\frac{\rho^{2}\gamma_{x}}{4\sigma_{x}^{4}}\cos\left(\left(\Omega+\delta\omega\right)t+\phi_{x}\right)-\frac{\rho^{2}m^{2}\gamma_{p}\omega^{2}}{4\sigma_{p}^{4}}\cos\left(\left(\Omega+\delta\omega\right)t+\phi_{p}\right),\label{eq:A6}
\end{equation}

\begin{equation}
A_{7}=\frac{\rho^{2}m^{2}\gamma_{p}\omega^{2}}{2\sigma_{p}^{4}}\cos\left(\Delta\omega-\phi_{p}\right)\label{eq:A7}
\end{equation}

and

\begin{equation}
A_{8}=\frac{\rho^{2}m^{2}\gamma_{p}\omega^{2}}{2\sigma_{p}^{4}}\cos\left(\left(\omega_{s}+\Omega\right)t+\phi_{p}\right).\label{eq:A8}
\end{equation}

\noindent The difference $\delta\omega$ sets the long period of the
fidelity, and results from the difference between the two Hamiltonians.
The frequency $2\omega\sim\omega_{s}$ is twice the frequency of rotation
around the fixed point at the origin. The overall coefficient corresponding
to the angular velocity is $\frac{\rho^{2}m^{2}\gamma_{p}\omega^{2}}{2\sigma_{p}^{4}}$
. 

\noindent The relations (\ref{eq:w_s like 2w}) - (\ref{eq:D_omega})
imply that (\ref{eq:s_x+s_p=00003Dsum}) with (\ref{eq:s_x+s_p=00003Dsum})
- (\ref{eq:A8}) oscillate with three different frequencies: $\omega_{s}$,
$\Delta\omega$ and $\delta\omega$, which are very different, and
satisfy $\omega_{s}\gg\Delta\omega\gg\delta\omega$. The corresponding
periods will be denoted by

\begin{equation}
T_{1}=\frac{2\pi}{\omega_{s}}\label{eq:T1 definition}
\end{equation}

\begin{equation}
T_{2}=\frac{2\pi}{\Delta\omega}\label{eq:T2 definition}
\end{equation}

\noindent and

\begin{equation}
T_{3}=\frac{2\pi}{\delta\omega}\label{eq:T3 definition}
\end{equation}

\noindent The frequencies $\omega_{s}$ and $\delta\omega$ (and consequently
$T_{1}$ and $T_{3}$), depend only on the classical frequencies $\omega_{1}$
and $\omega_{2}$. The frequency $\Delta\omega$ depends on $\Omega$
that is not related to any of the classical frequencies. 

Note that also harmonics of these three basic frequencies may be present.
Since this is a very heuristic theory, also deviations and splitting
of the frequency peaks are expected.

\section{Fidelity oscillations\label{sec:Fidelity-oscillations}}

In this section we present oscillations of the fidelity. In previous
work \cite{Yevgeny_paper}, the fidelity oscillations in absence of
interactions were calculated. 

In particular, there were identified two frequencies. These frequencies
are of pure classical origin. One of them denoted by $\omega_{s}$
,is related to the classical motion around the elliptic fixed point.
The second frequency is $\delta\omega$. In presence of interactions
an intermediate frequency $\omega_{I}$ is found. In this section
we report the numerical values of these frequencies for various values
of parameters. 

\begin{figure}[H]
\centering{}\subfloat[\label{fig:Fidelity vs. kicks}]{\begin{centering}
\includegraphics[scale=0.6]{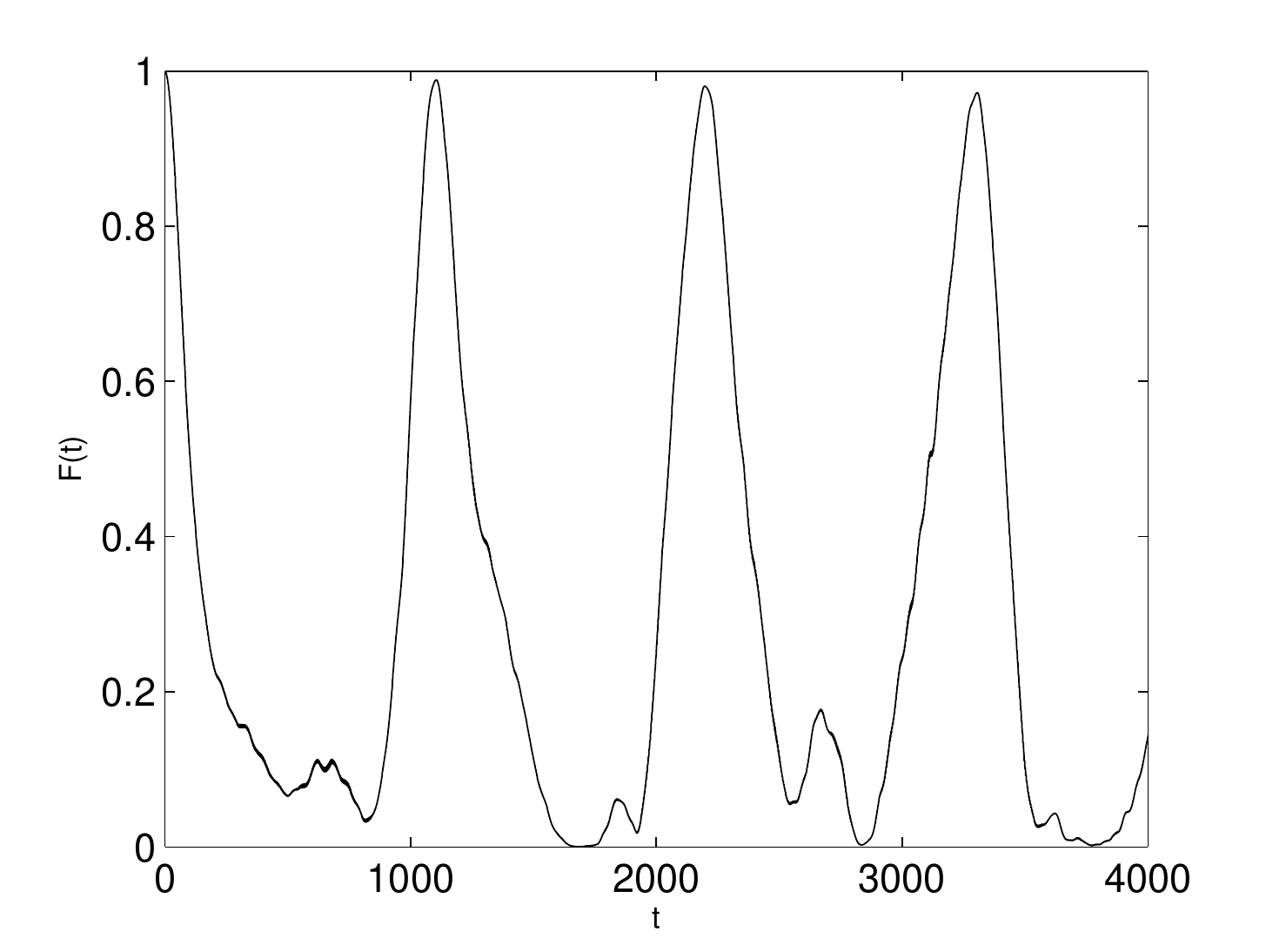}
\par\end{centering}

}\hfill{}\subfloat[\label{fig:Fidelity Fourier transform log10}]{\begin{centering}
\includegraphics[scale=0.6]{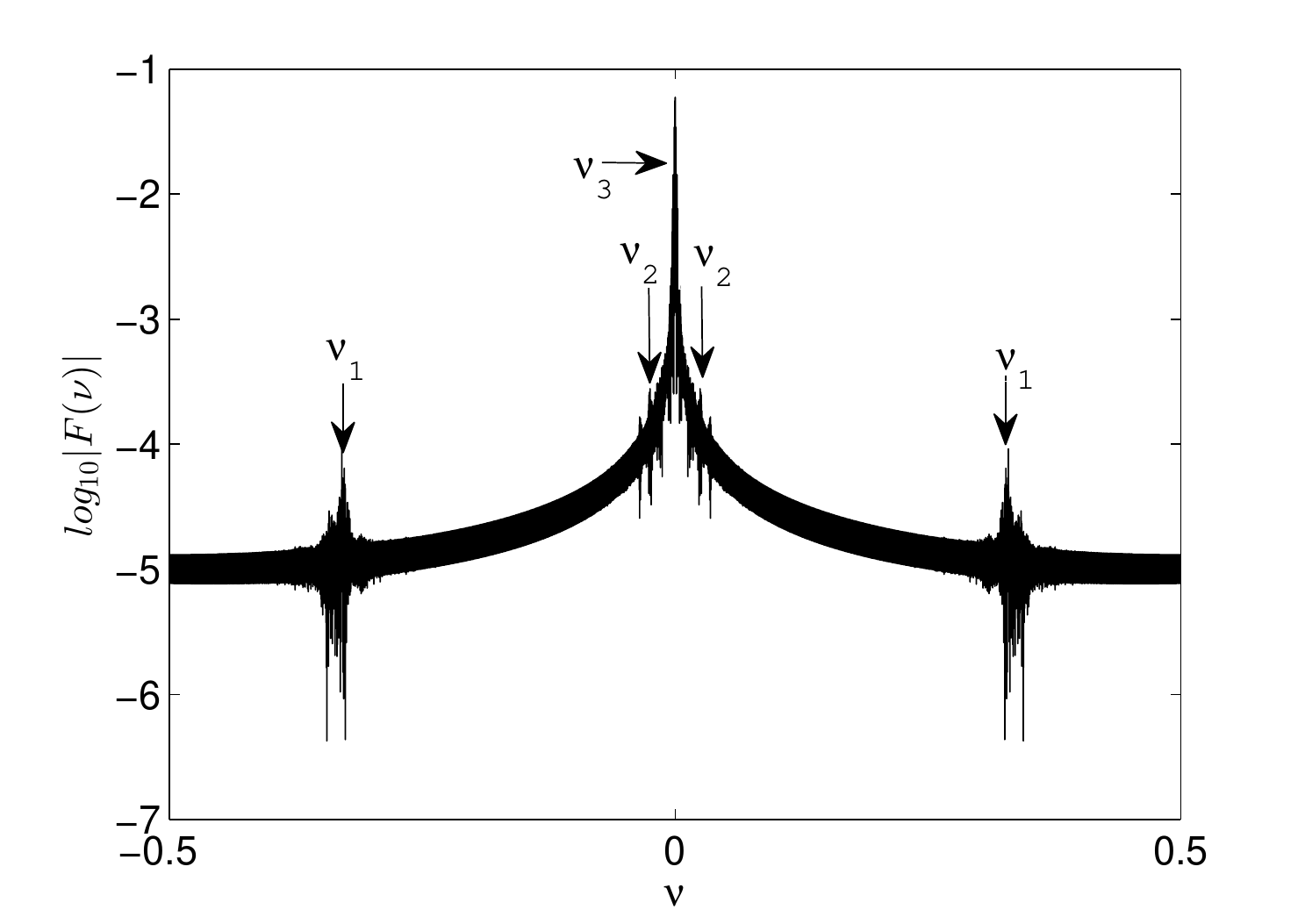}
\par\end{centering}

}\caption{\label{fig:Fidelity vs. kicks. (0.3,0)}The fidelity\textbf{ }for
\textbf{$\left(x_{0}\left(t=0\right),p_{0}\left(t=0\right)\right)=\left(0.18,0\right)$,
$\beta=6\cdot10^{-5}$ }and\textbf{ $\tau=0.01$} (a): as function
of number of kicks\textbf{ }(b):\textbf{ $log_{10}|\hat{F}\left(\nu\right)|$
}as a function of the inverse number of kicks\textbf{ $\nu$}. }
\end{figure}

In all calculations presented here we used two Hamiltonians of the
form (\ref{eq:Hamiltonian with interactions}) with the values of
the stochasticity parameter $K$ that takes values that are close,
namely, $K_{1}=1$ and $K_{2}=1.01$. We launched an initial wave
packet of the form (\ref{eq: Coherent state wave function}) for various
initial values of $x_{0}\left(t=0\right)$ and $p_{0}\left(t=0\right)$.
Each wave packet was iterated using the one step propagator (\ref{eq:One kick propagator}).
The fidelity was calculated from (\ref{eq:Fidelity bra ket form}).
Plots of the form Fig. \ref{fig:Fidelity vs. kicks} with the corresponding
Fourier transform in Fig. \ref{fig:Fidelity Fourier transform log10}
were calculated from

\noindent 
\begin{equation}
\hat{F}\left(\nu\right)=\intop_{-\infty}^{\infty}F\left(t\right)e^{-i2\pi\nu t}dt\label{eq:Fourier transform general form}
\end{equation}
were generated. The dominant frequencies are marked by arrows in Fig.
\ref{fig:Fidelity Fourier transform log10}. We repeated the calculation
for different initial values of $x_{0}\left(t=0\right)$, $p_{0}\left(t=0\right)$
and $\beta$. 

\noindent In Fig. \ref{fig:Fidelity vs. kicks. (0.3,0)} we found
numerically that the fidelity exhibits three frequencies. A large
frequency $\nu_{1}\sim0.33\left[kicks^{-1}\right]$, corresponding
to period $T_{1}\sim3\left[kicks\right]$, an intermediate frequency,
$\nu_{2}\sim0.025\left[kicks^{-1}\right]$, corresponding to $T_{2}\sim40\left[kicks\right]$,
and a small frequency $\nu_{3}\sim0.001\left[kicks^{-1}\right]$,
corresponding to $T_{3}\sim1000$$\left[kicks\right]$. These results
were repeated for various initial values of $x_{0}\left(t=0\right)$,
$p_{0}\left(t=0\right)$ and $\beta$ and are presented in Figs. \ref{fig:T1,T2 as function of beta, (0.18,0)}
- \ref{fig:T1,T2,T3 vs. x0}. In Fig. \ref{fig:T1,T2 as function of beta, (0.18,0)},
the periods $T_{1}$, $T_{2}$ and $T_{3}$ found from plots similar
to the ones presented in Fig. \ref{fig:Fidelity vs. kicks. (0.3,0)},
are plotted as a function of $\beta$ for $\left(x_{0}\left(t=0\right),p_{0}\left(t=0\right)\right)=\left(0.18,0\right)$
and $\tau=0.01$. Similar results are found for various initial conditions
such that $x_{0}\left(t=0\right)\geq0.14$, $p_{0}\left(t=0\right)=0$
and $\tau=0.01$. Note that $T_{1}$ is slowly increasing with $\beta$. 

In Fig. \ref{fig:T1,T2,T3 vs. x0}, the periods $T_{1},T_{2}$ and
$T_{3}$ as function of $x_{0}\left(t=0\right)$ are presented for
$p_{0}\left(t=0\right)=0$, $\beta=6\cdot10^{-5}$ and $\tau=0.01$.
The results for $x_{0}\left(t=0\right)=0$, $p_{0}\left(t=0\right)\neq0$
are similar. 

\begin{figure}[H]
\centering{}\subfloat[\label{fig:T1 vs. beta (0.18,0)}]{\begin{centering}
\includegraphics[scale=0.57]{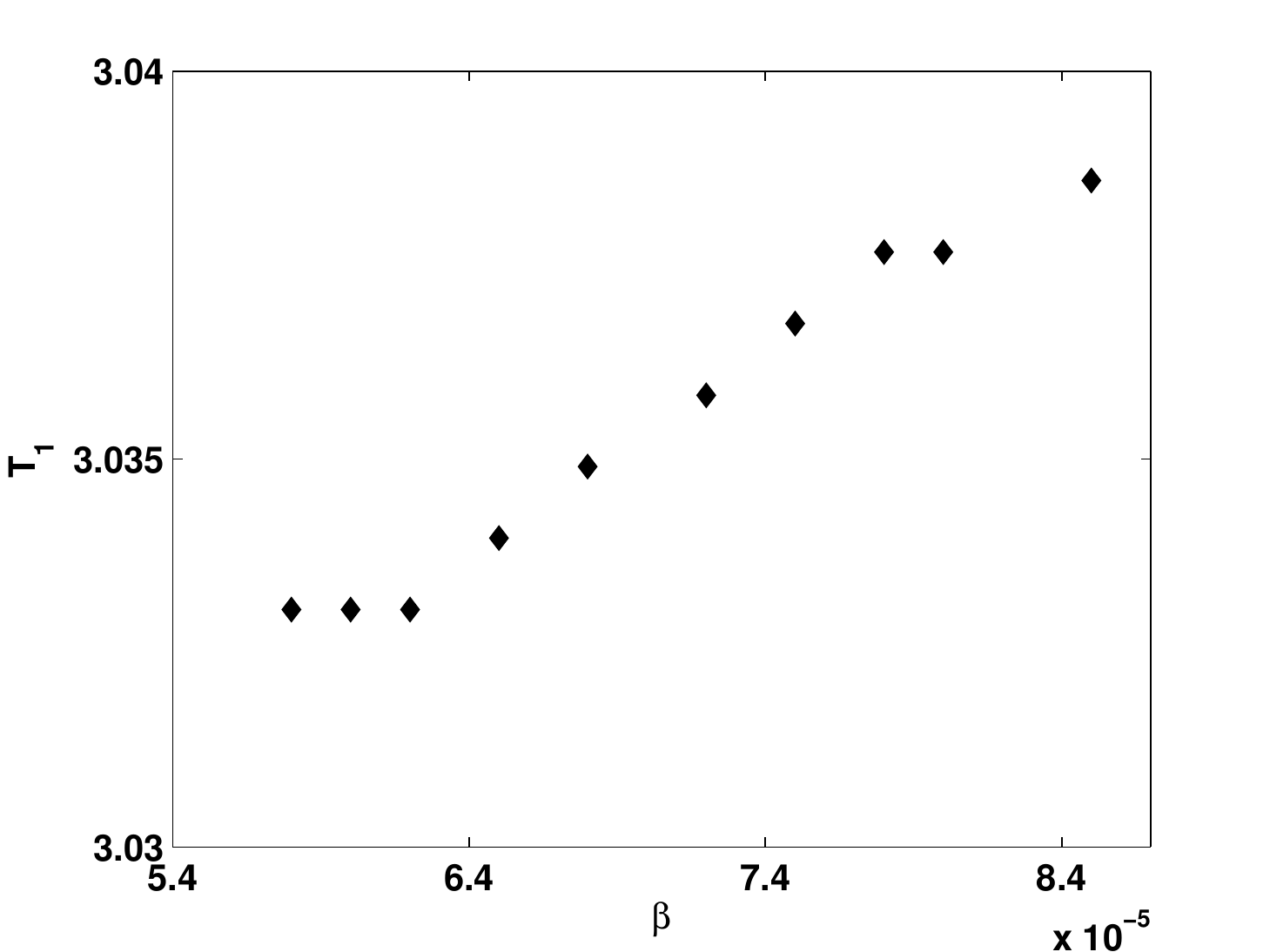}
\par\end{centering}

}\hfill{}\subfloat[\label{fig:T2 vs. beta, (0.18,0)}]{\begin{centering}
\includegraphics[scale=0.57]{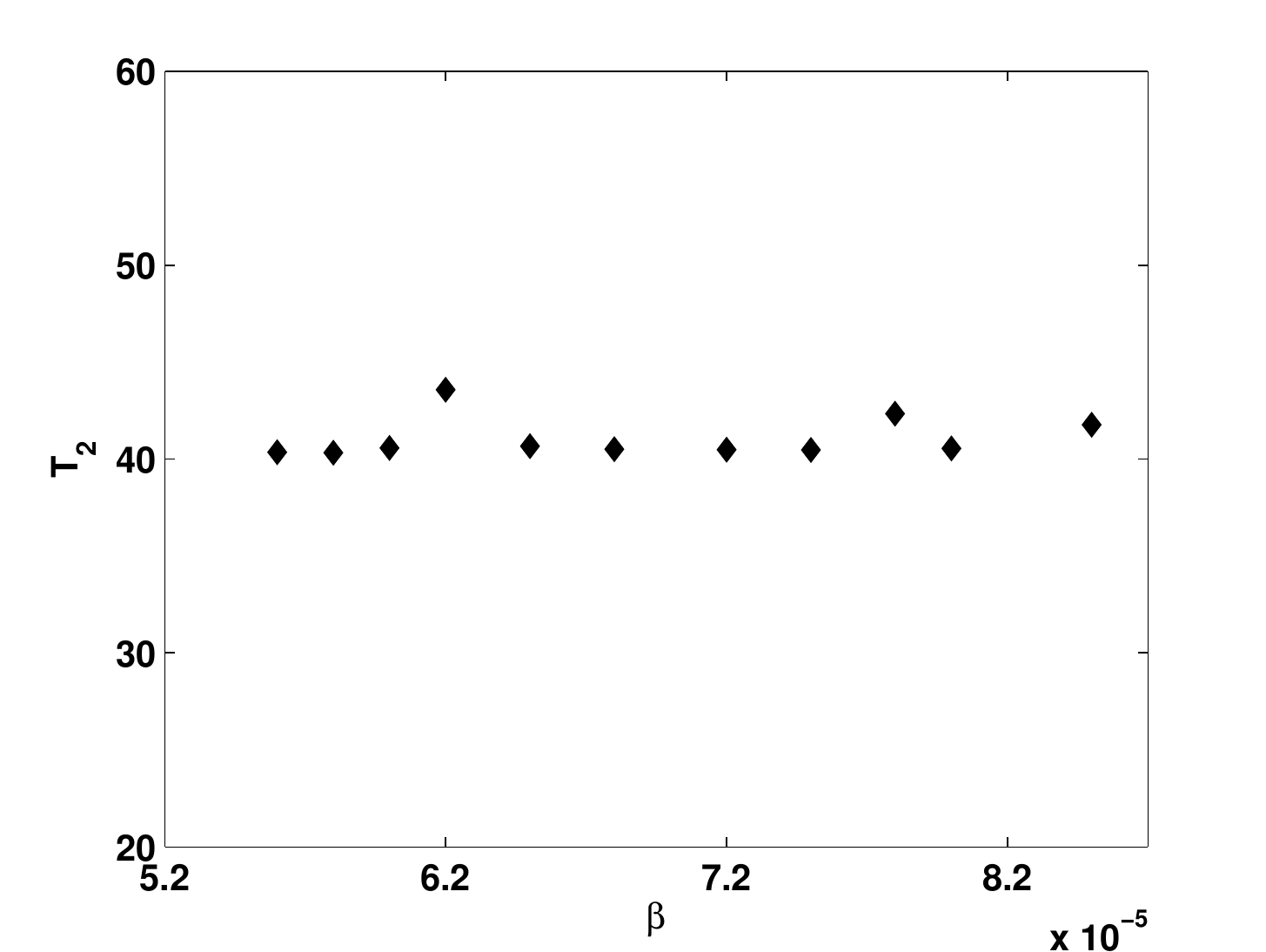}
\par\end{centering}

}\hfill{}\subfloat[\label{fig:T3 vs. beta, (0.18,0)}]{\centering{}\includegraphics[scale=0.57]{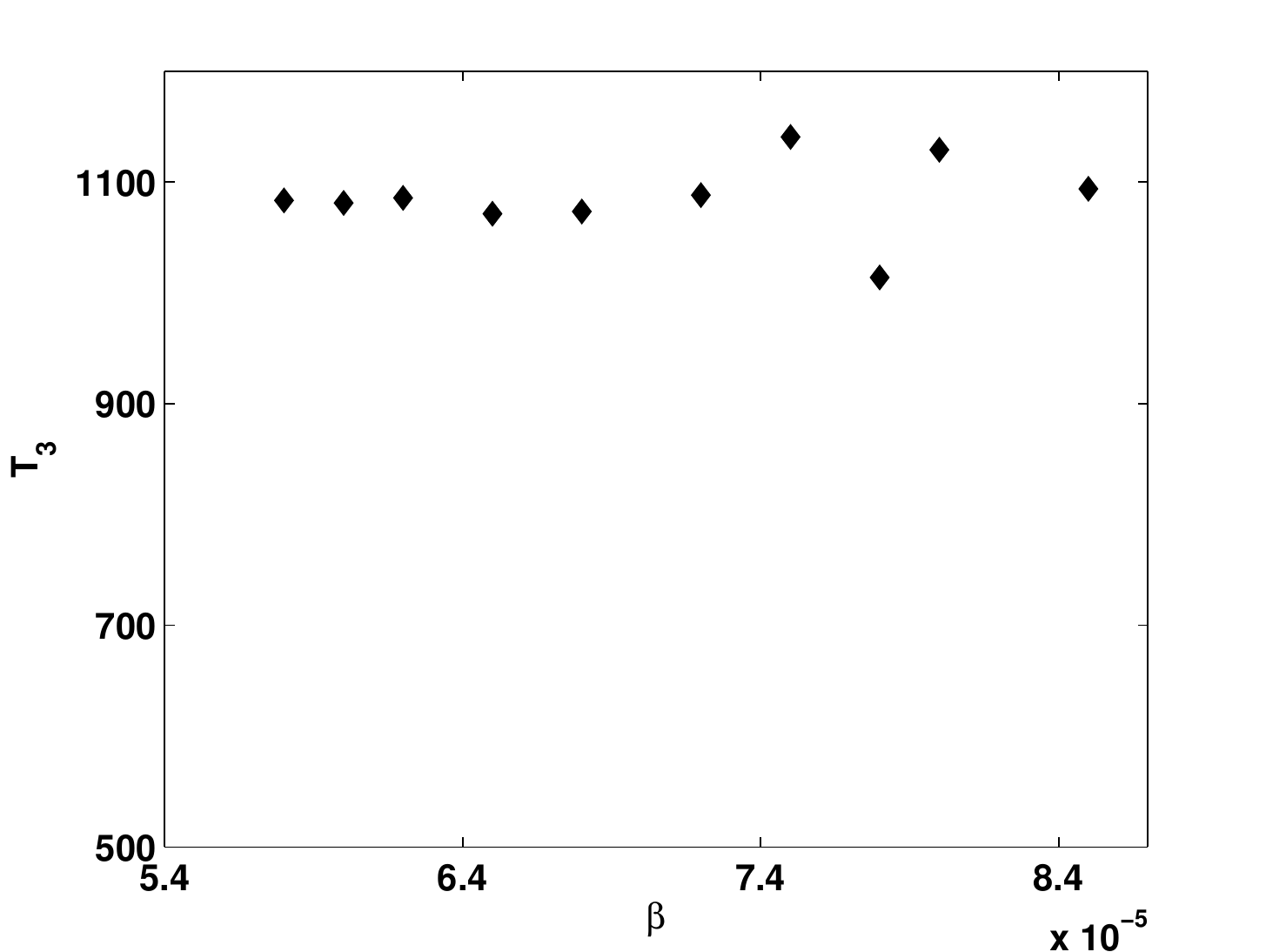}}\caption{\label{fig:T1,T2 as function of beta, (0.18,0)}Various periods of
the fidelity as a function of $\beta$ for $\left(x_{0}\left(t=0\right),p_{0}\left(t=0\right)\right)=\left(0.18,0\right)$
and $\tau=0.01$ (a): \textbf{$T_{1}$} (b): $T_{2}$ (c): $T_{3}$}
\end{figure}

\begin{center}
\begin{figure}[H]
\centering{}\subfloat[\label{fig:T1 vs. x0}]{\centering{}\includegraphics[clip,scale=0.57]{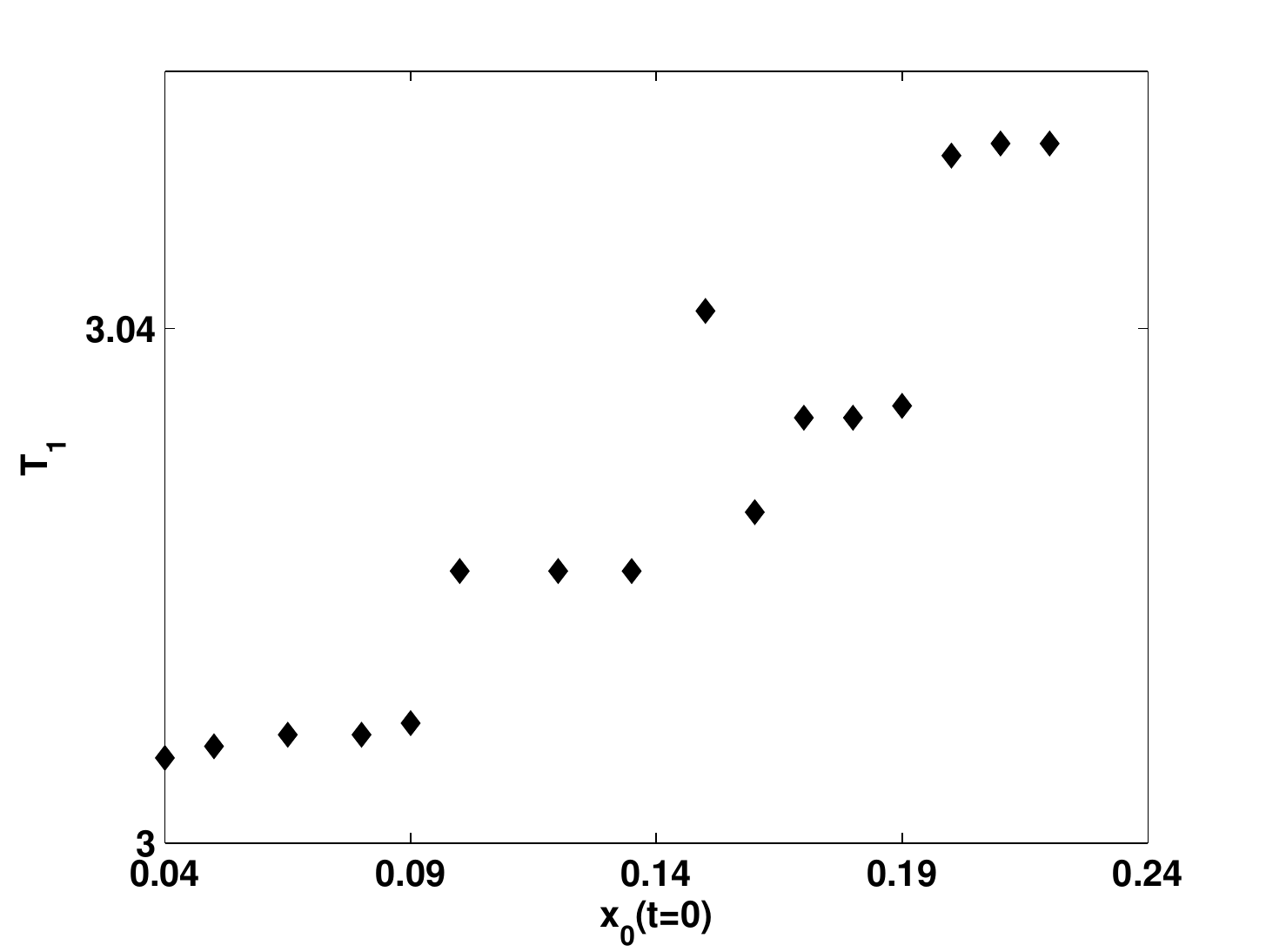}}\hfill{}\subfloat[\label{fig:T2 vs. x0}]{\centering{}\includegraphics[scale=0.57]{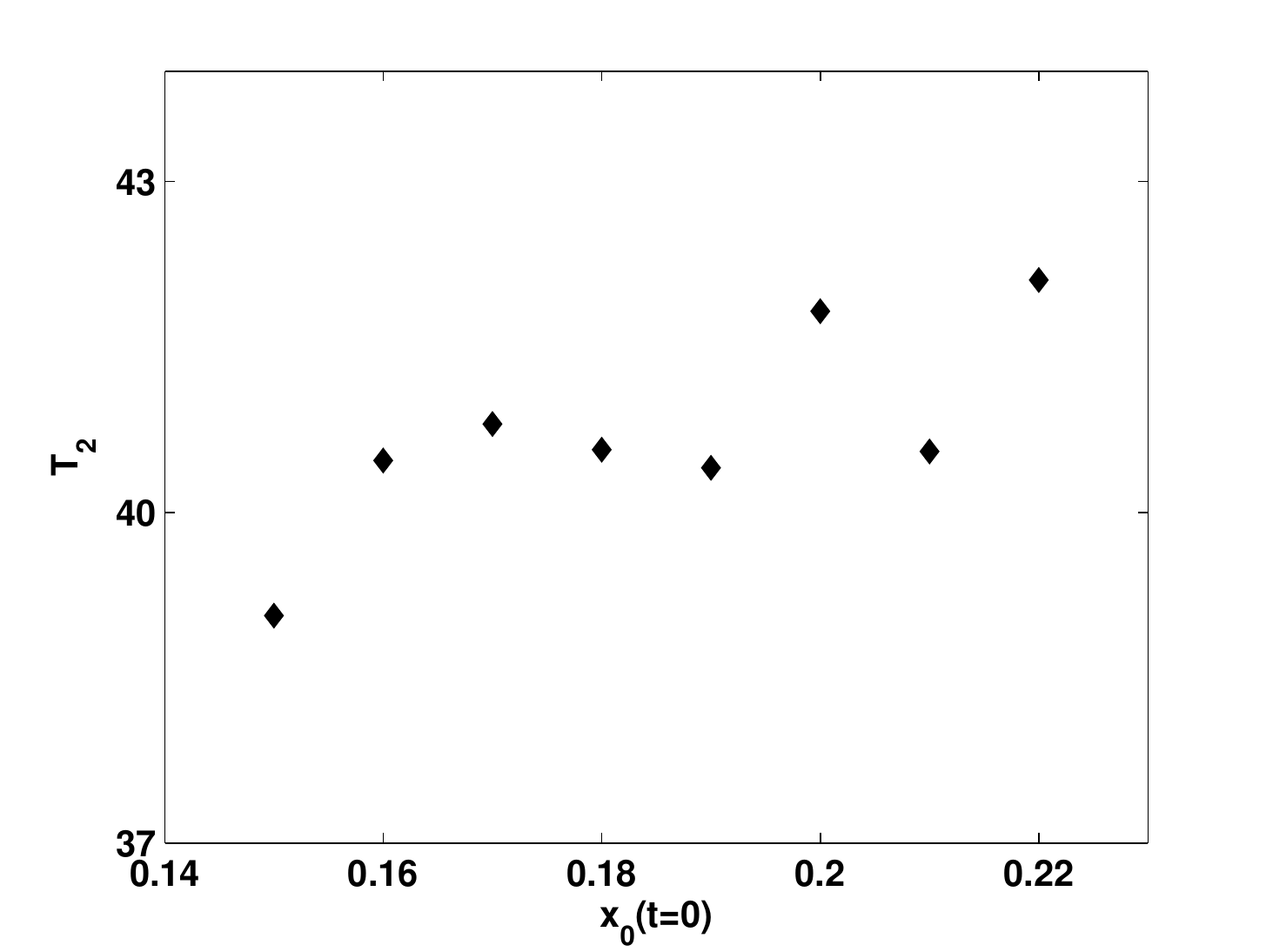}}\hfill{}\subfloat[\label{fig:T3 vs. x0}]{\centering{}\includegraphics[scale=0.57]{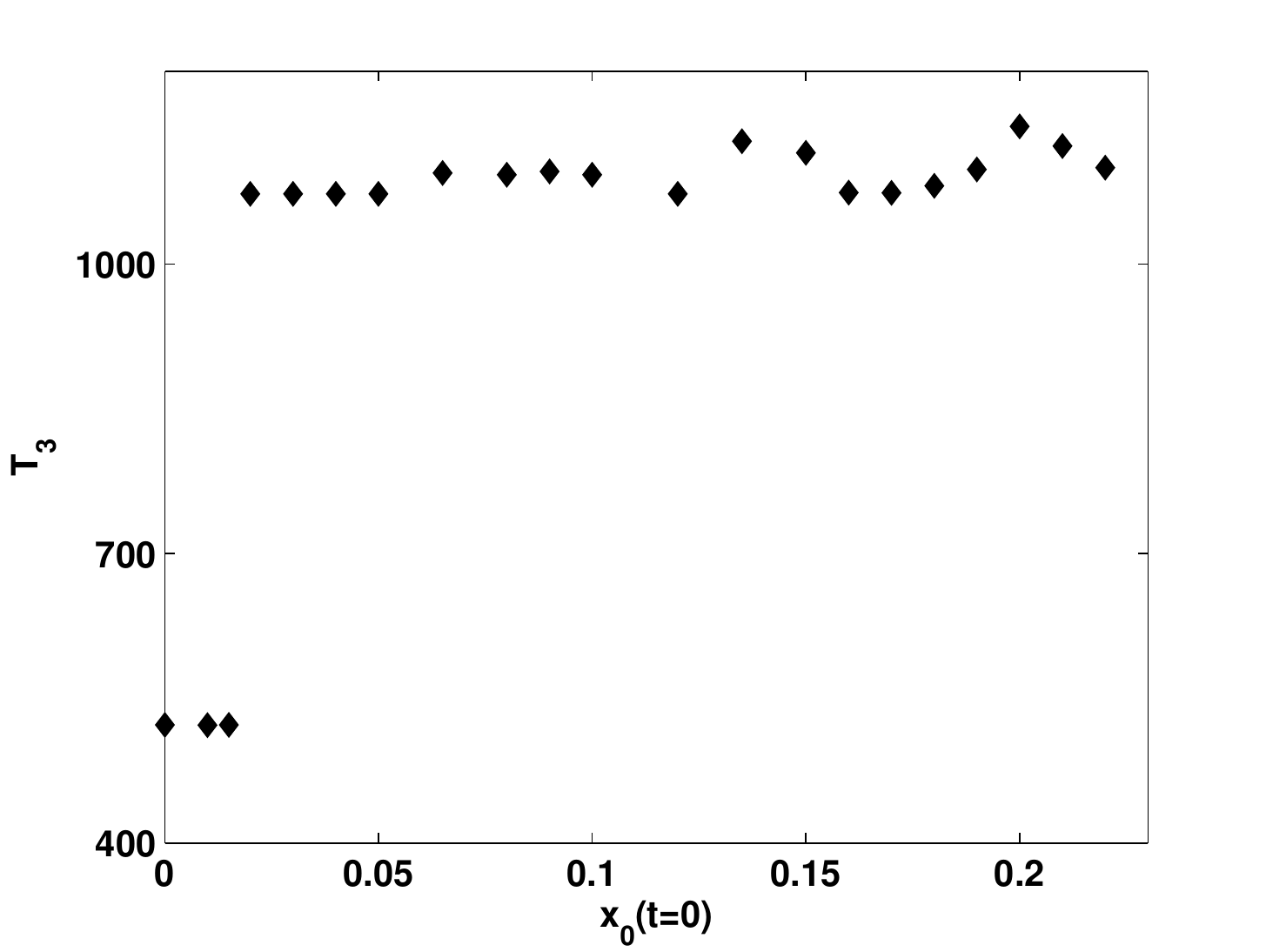}}\caption{\label{fig:T1,T2,T3 vs. x0}The various periods as a function of $x_{0}\left(t=0\right)$
for $p_{0}\left(t=0\right)=0$, $\beta=6\cdot10^{-5}$ and $\tau=0.01$
(a): $T_{1}$ (b): $T_{2}$ (c): $T_{3}$ }
\end{figure}

\par\end{center}

In all situations we found that the period $T_{1}$ varies between
$3$ and $3.05$ kicks. It is very close to the value $T_{1}=\frac{2\pi}{\omega_{s}}\simeq\frac{\pi}{\omega_{1}}=3$
kicks, where $\omega_{1}$ is given by (\ref{eq:Explicit omega_1}).
The period is systematically increasing with $x_{0}\left(t=0\right)$
and $p_{0}\left(t=0\right)$ (see Fig. \ref{fig:T1 vs. x0}). The
reason is the deviation of the frequency from the value found in the
vicinity of the fixed point at the origin. This can be verified by
direct iteration of the map (\ref{eq:Classical map for momentum})
- (\ref{eq:Classical map for position}). 

For $x_{0}\left(t=0\right)$ that is sufficiently large, the period
$T_{3}$ was found to take the value $T_{3}\sim1100$ kicks. It is
close to value predicted from pure classical dynamics without interactions,
$T_{3}=1091.8$ kicks for $K_{1}=1$ and $K_{2}=1.01$, calculated
using (\ref{eq:T3 definition}). For $x_{0}\left(t=0\right)=p_{0}\left(t=0\right)=0$,
we expect that $T_{3}=\frac{\pi}{\delta\omega}$, rather than $\frac{2\pi}{\delta\omega}.$
This is because of the symmetry of the initial condition. Each point
of a trajectory generated by $H_{1}$ is chasing a point generated
by $H_{2}$ which is its reflection through the origin of the phase
space and, therefore, is found first at an angle of $\pi$ and not
$2\pi$. Indeed, this was found for sufficiently small $x_{0}\left(t=0\right)$
(see Fig. \ref{fig:T3 vs. x0}). 

In summary, the periods $T_{1}$ and $T_{3}$ are of pure classical
origin. These were found in \cite{Yevgeny_paper}. Here we found that
these are weakly affected by the interactions. The intermediate period
$T_{2}$ is found to be $T_{2}\sim40$ kicks (see Figs. \ref{fig:T2 vs. beta, (0.18,0)}
and \ref{fig:T2 vs. x0}). This period was not found in absence of
interactions. 

We turn now to the exploration of the origin of this new period.

\section{The origins of the intermediate period\label{sec:The-origins-of_intermediate_period}}

In this section we will demonstrate that the intermediate period results
from the oscillation of the width of the wavefunction. 

\noindent The Fourier transform of the width (\ref{eq:width of wave packet, general expression})

\begin{equation}
\hat{f}_{\Delta}\left(\nu\right)=\intop_{-\infty}^{\infty}\left\langle \Delta x\left(t\right)\right\rangle ^{2}e^{-i2\pi\nu t}dt\label{eq:Fourier transform of width}
\end{equation}

\noindent was computed for $\psi$ which was derived from an initially
coherent state of the form (\ref{eq: Coherent state wave function})
by application of the evolution operator (\ref{eq:One kick propagator}).
We found that it exhibits the peaks $T_{1}=3.05\left[kicks\right],$
$\nu_{1}=0.328\left[\frac{1}{kick}\right]$, $\left(\omega_{1}\simeq\omega=2.06\left[\frac{1}{kick}\right]\right),$
$T_{width}=3.24$$\left[kicks\right]$ , $\nu_{width}=0.309\left[\frac{1}{kick}\right]$
and $\Omega_{width}=1.94\mbox{\ensuremath{\left[\frac{1}{kick}\right]}}$
for $\beta=6\cdot10^{-5}$, $\tau=0.01$ , $\left(x_{0}\left(t=0\right),p_{0}\left(t=0\right)\right)=\left(0,0.14\right)$
and for $\left(x_{0}\left(t=0\right),p_{0}\left(t=0\right)\right)=\left(0.18,0\right)$. 

\begin{figure}[H]
\begin{centering}
\includegraphics[scale=0.55]{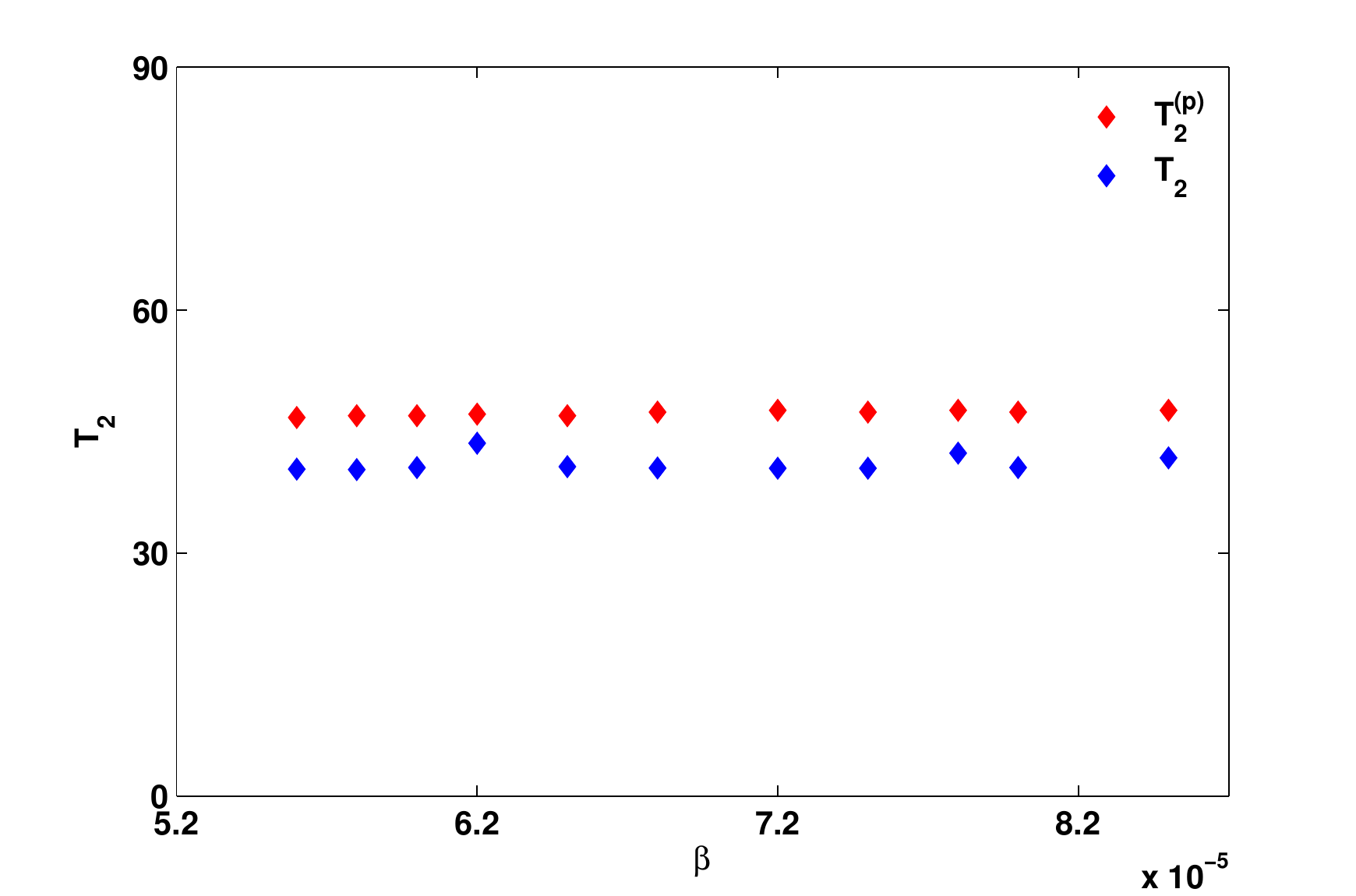}
\par\end{centering}

\caption{\label{fig:Intermediate period - comparison with calculation}(Color
online) The predicted intermediate period of the fidelity\textbf{
$T_{2}^{(p)}$} compared to $T_{2}$, found directly from the Fourier
transform of the fidelity, as function of $\beta$, for $\left(x_{0}\left(t=0\right),p_{0}\left(t=0\right)\right)=\left(0.18,0\right)$\textbf{
}and\textbf{ $\tau=0.01$ .}}
\end{figure}

We note that indeed $\Omega_{width}$ which was found numerically
is close to $2\omega$. Taking $\Omega=\Omega_{width}$, we use (\ref{eq:D_omega})
to calculate 

\begin{equation}
\Delta\omega=2\omega-\Omega_{width},
\end{equation}

\noindent and find the predicted intermediate period $T_{2}^{(p)}=\frac{2\pi}{\Delta\omega}$,
where $\Omega_{width}$ is found from the numerical calculations of
(\ref{eq:Fourier transform of width}). Comparison between this value
and $T_{2}$ calculated from the fidelity Fourier transform (\ref{eq:Fourier transform general form})
is shown in Fig. \ref{fig:Intermediate period - comparison with calculation}.
The difference is small, as expected from section \ref{sub:Fidelity for coherent states with weak interactions}. 

\begin{figure}
\begin{centering}
\includegraphics[scale=0.6]{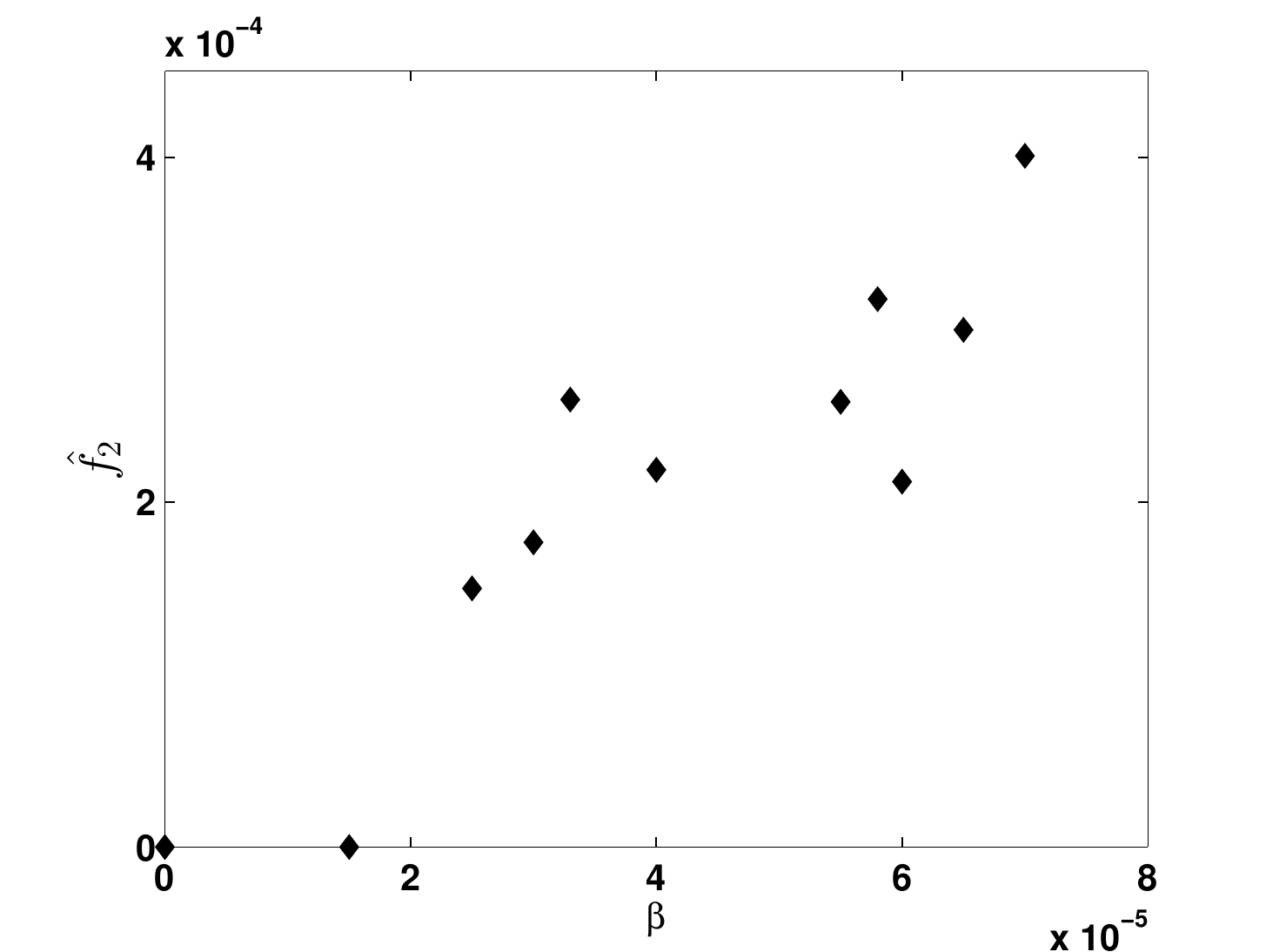}
\par\end{centering}

\caption{\label{fig:Amplitude vs. beta}Fidelity Fourier transform amplitude
of $f_{2}$ as function of $\beta$ for $\tau=0.01$, $\left(x_{0}\left(t=0\right),p_{0}\left(t=0\right)\right)=\left(0.18,0\right)$
and $7000$ kicks }
\end{figure}

An obvious question is how is it possible that an oscillation with
the same period $T_{2}$ is found for all $\beta\neq0$, but no such
oscillation is found for $\beta=0$. For this purpose, the amplitude
of this oscillation Fourier transform peak $\hat{f_{2}}$ as function
of $\beta$ is plotted in Fig. \ref{fig:Amplitude vs. beta}. It can
be seen that the amplitude decreases as $\beta$ decreases. 

The generation of the intermediate period is not characteristic of
the fidelity but will show up in any correlation function involving
overlap of Wigner functions. The fidelity is the overlap of Wigner
functions at the same time but different Hamiltonians. Similar behavior
is found for overlap of the Wigner functions for the same Hamiltonian
but at different times $n$ and $n-\Delta n$ defined by 

\begin{equation}
G\left(n\right)=\iintop_{-\infty}^{\infty}W_{n}\left(x,p\right)W_{n-\Delta n}\left(x,p\right)dxdp\label{eq:Wigner correlation integral}
\end{equation}

\noindent and calculated in detail in Appendix B.

\begin{figure}[H]
\centering{}\subfloat[\label{fig:Wigner 990-995 Beta =00003D 0}]{\begin{centering}
\centering\includegraphics[scale=0.58]{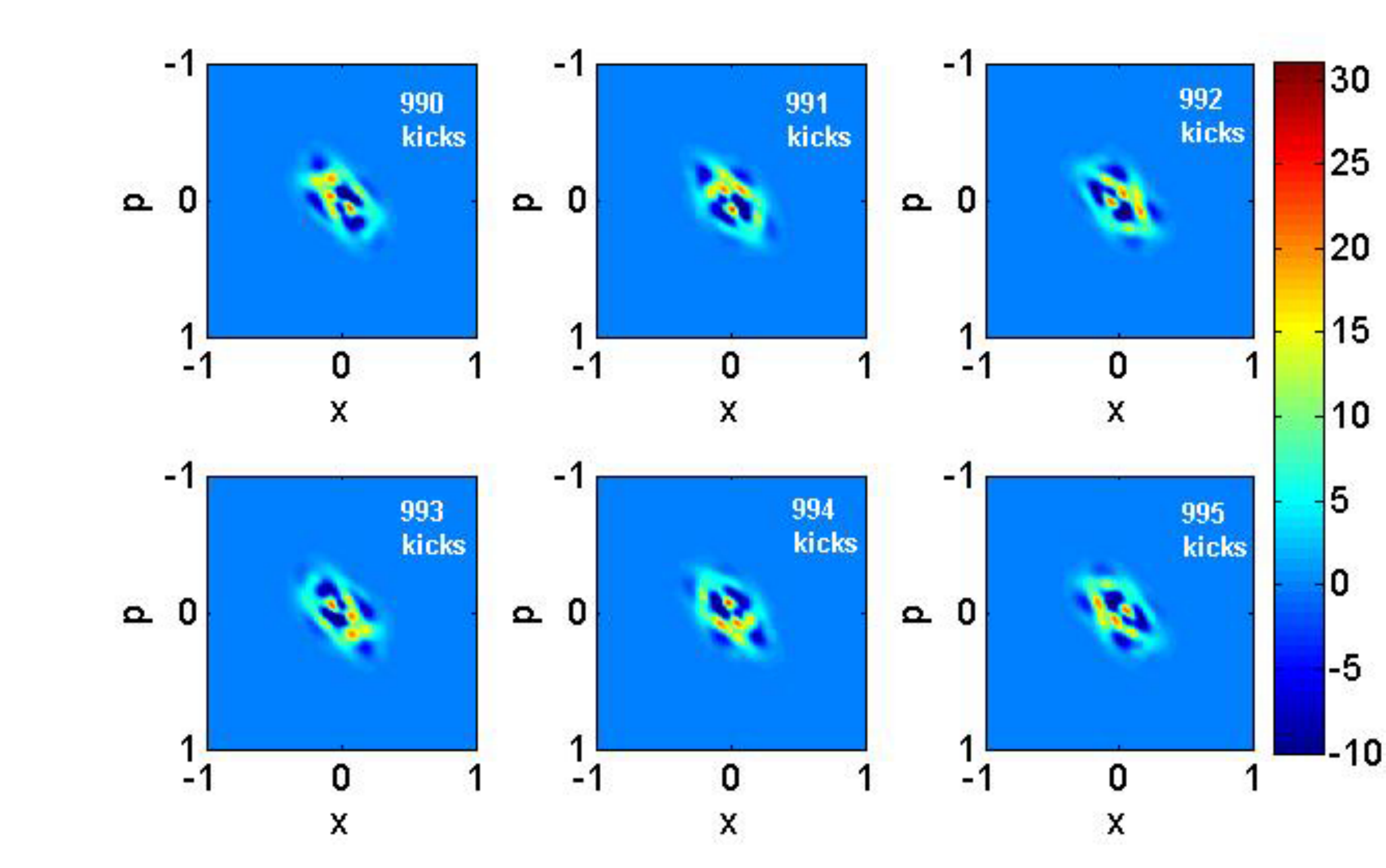}
\par\end{centering}

}\hfill{}\subfloat[\label{fig:Wigner 990-995 Beta=00003D6e-5}]{\begin{centering}
\centering\includegraphics[scale=0.55]{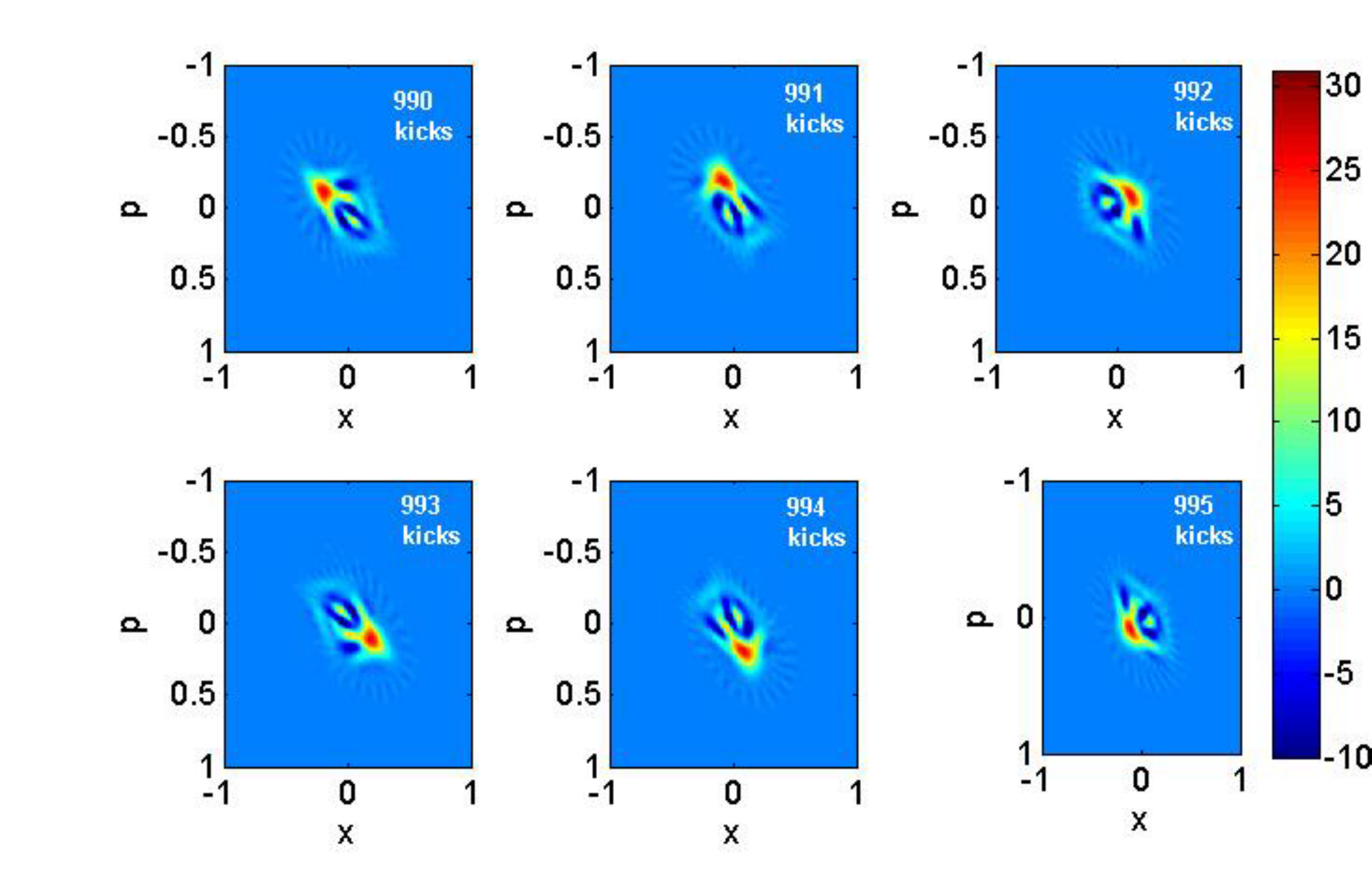}
\par\end{centering}

}\caption{\textbf{\label{fig:Wigner function at different times-1K kicks}}(Color
online) The Wigner function for 990 - 995 kicks for \textbf{$\left(x_{0},p_{0}\right)=\left(0.18,0\right)$,
$\beta=6\cdot10^{-5}$} and\textbf{ $\tau=0.01.$ }(a): $\beta=$0,
(b): $\beta=6\cdot10^{-5}$.}
\end{figure}

First, we note that the Wigner function rotates around the elliptic
point as demonstrated in Fig. \ref{fig:Wigner function at different times-1K kicks}
for 990-995 kicks. In Fig. \ref{fig:Wigner 990-995 Beta =00003D 0}
we see that for $\beta=0$ the function shape is smeared over the
phase space. In Fig. \ref{fig:Wigner 990-995 Beta=00003D6e-5} we
see that the Wigner function for $\beta\neq0$ is localized due to
the interactions. Hence, in this case interactions tend to localize
the Wigner function in phase space. In presence of interactions the
general form is indeed (\ref{eq:Wigner function, harm. osc. general form})
with (\ref{eq:width in x with oscillations}) and (\ref{eq:width in p with oscillations}),
replacing $\sigma_{x_{i}}^{2}$ and $\sigma_{p_{i}}^{2}$, respectively. 

The Fourier transform of the correlation function $G\left(n\right)$
and the Fourier transform of the fidelity with the same parameters
as in Fig. \ref{fig:Fidelity vs. kicks. (0.3,0)} have been compared.
The intermediate frequency found from the fidelity is $\nu_{2}=0.025\left[kicks^{-1}\right]$
, $\omega_{2}=0.157\left[\frac{1}{kick}\right]$ with period $T_{2}=40.57\left[kicks\right]$,
and the frequency found from $G\left(n\right)$ is $\nu_{2}=0.024\left[kicks^{-1}\right]$
with $\omega_{2}=0.151\left[\frac{1}{kick}\right]$ and $T_{2}=42.52\left[kicks\right]$.
The results are similar in both cases. 

\section{Summary and discussion}

The effects of weak inter-particle interactions on the quantum fidelity
were calculated for kicked particles. The calculation was performed
for a specific model where the interaction was introduced during the
kicks. The results were found to be qualitatively similar to the ones
found where the interactions were introduced between kicks \cite{Mark_Update}.
We found that the periods that were obtained in absence of the interactions,
namely, $T_{1}$ and $T_{3}$, are found also in the presence of the
interactions. In presence of the interactions, another period, namely,
$T_{2}$, was found. We explored the mechanism of the generation of
this frequency. It results of the interplay of the oscillations of
the width of the wave function (or Wigner function) in phase space
and the rotation of its center around the elliptic fixed point. It
is $\Delta\omega$ of (\ref{eq:D_omega}) that was derived in the
framework of the heuristic model outlined in Sec. \ref{sec:Fidelity for weak interactions}
and tested numerically in Sec. \ref{sec:The-origins-of_intermediate_period}.
In Fig. \ref{fig:Wigner function at different times-1K kicks} it
was verified that the heuristic picture of Sec. \ref{sec:Fidelity for weak interactions}
holds for the kicked model presented in the introduction. The frequencies
found in this work for the fidelity are found also for other correlation
functions of Wigner functions. 

In this work we focused on dynamics of wave packets in the vicinity
of the elliptic fixed point $\left(x,p\right)=\left(0,0\right)$ for
the classical phase portrait shown in Fig. \ref{fig:Phase space for K=00003D1}. 

The existence of the intermediate frequency $\Delta\omega$ ((\ref{eq:D_omega}))
and its origin is the main result of this paper. From the analysis
of \cite{Mark_paper} it is plausible that the origin of this frequency
is semiclassical. The meaning is that the term $\beta\left|\psi\left(x\right)\right|^{2}$
in (\ref{eq:Hamiltonian with interaction replacement}), (\ref{eq:One kick propagator})
and (\ref{eq:Hamiltonian with interactions}) acts as a potential.
The intermediate frequency is not found numerically if the center
of the wave packet is too close to the elliptic fixed point. A possible
explanation is that in such a situation one does not have the possibility
to separate the rotations of the center of the packet and the oscillation
of the width. 

As one increases the distance of the wave packet from the fixed point
at the origin while keeping the nonlinearity fixed, the variation
of the rotation frequency increases and the packet spreads over a
ring in phase space, as is the case for $\beta=0$ (see Fig. \ref{fig:Wigner 990-995 Beta =00003D 0}).
In such a situation the picture of Sec. \ref{sec:Fidelity for weak interactions}
is violated. Nevertheless, the same intermediate frequency $\Delta\omega$
is numerically found. This should be left for further studies.

\section*{Acknowledgements}

It is our great pleasure to thank Dr. Mark Herrera for illuminating,
inspiring and critical discussions and communications. This work was
partly supported by the Israel Science Foundation (ISF), Grant number
1028/12 , by the US-Israel Binational Science Foundation (BSF), Grant
number 2010132 , by the Minerva Center of Nonlinear Physics of Complex
Systems, by the New York Metropolitan Fund and by the Shlomo Kaplansky
academic chair. 

\appendix

\section{Fidelity calculation details}

The equations (\ref{eq:s_x with oscillations}) and (\ref{eq:s_p with oscillations})
are simplified by means of a Taylor expansion 

\begin{equation}
\frac{1}{1-x}\simeq1+x+O\left(x^{2}\right),\label{eq:Taylor expansion}
\end{equation}

\noindent combined with

\begin{equation}
\cos\alpha\cdot\cos\beta=\frac{1}{2}\left(\cos\left(\alpha-\beta\right)+\cos\left(\alpha+\beta\right)\right),\label{eq:cosine trig formula}
\end{equation}

\noindent the equations take the form

\begin{eqnarray}
s_{x} & = & \frac{\rho^{2}}{2\sigma_{x}^{2}}-\frac{\rho^{2}}{2\sigma_{x}^{2}}\cos\left(\delta\omega\cdot t\right)+\frac{\rho^{2}\nu}{4\sigma_{x}^{4}}\cos\left(\left(\Omega-\delta\omega\right)t+\phi_{x}\right)\\
 &  & +\frac{\rho^{2}\nu}{4\sigma_{x}^{4}}cos\left(\left(\Omega+\delta\omega\right)t+\phi_{x}\right)-\frac{\rho^{2}\nu}{2\sigma_{x}^{4}}\cos\left(\Omega\cdot t+\phi_{x}\right)\nonumber 
\end{eqnarray}

\noindent and

\begin{eqnarray}
s_{p} & = & \frac{\rho^{2}m^{2}\omega^{2}}{2\sigma_{p}^{2}}-\frac{\rho^{2}m^{2}\omega^{2}}{\sigma_{p}^{2}}\cos\left(2\omega\cdot t\right)+\frac{\rho^{2}m^{2}\omega^{2}}{2\sigma_{p}^{2}}\cos\left(\delta\omega\cdot t\right)+\\
 &  & +\frac{\rho^{2}m^{2}\gamma\omega^{2}}{2\sigma_{p}^{4}}\cos\left(\left(2\omega+\Omega\right)t+\phi_{p}\right)-\frac{\rho^{2}m^{2}\gamma\omega^{2}}{4\sigma_{p}^{4}}\cos\left(\left(\Omega-\delta\omega\right)t+\phi_{p}\right)\nonumber \\
 &  & -\frac{\rho^{2}m^{2}\gamma\omega^{2}}{4\sigma_{p}^{4}}\cos\left(\left(\Omega+\delta\omega\right)t+\phi_{p}\right)+\frac{\rho^{2}m^{2}\gamma\omega^{2}}{2\sigma_{p}^{4}}\cos\left(\left(2\omega+\Omega\right)t+\phi_{p}\right)\nonumber \\
 &  & -\frac{\rho^{2}m^{2}\gamma\omega^{2}}{4\sigma_{p}^{4}}\cos\left(\left(\Omega-\delta\omega\right)t+\phi_{p}\right)-\frac{\rho^{2}m^{2}\gamma\omega^{2}}{4\sigma_{p}^{4}}\cos\left(\left(\Omega+\delta\omega\right)t+\phi_{p}\right)+\nonumber \\
 &  & +\frac{\rho^{2}m^{2}\gamma\omega^{2}}{2\sigma_{p}^{4}}\cos\left(\Delta\omega-\phi_{p}\right)-\frac{\rho^{2}m^{2}\gamma\omega^{2}}{2\sigma_{p}^{4}}\cos\left(\Omega\cdot t+\phi_{p}\right).\nonumber 
\end{eqnarray}

\noindent From this, one finds (\ref{eq:s_x+s_p=00003Dsum}) - (\ref{eq:A8}).

\section{Correlation of the Wigner function at various times}

In this Appendix we identify the frequencies of $G\left(n\right)$
defined by (\ref{eq:Wigner correlation integral}), where $\Delta n$
is fixed. The derivation is similar to the derivation of the fidelity
oscillations is Sec. \ref{sec:Fidelity for weak interactions} and
Appendix A. First, we assume there are no interactions and then we
add the effect of weak interactions. We consider wave packets near
the elliptic fixed point $\left(x,p\right)=\left(0,0\right)$ and
as in the case of the fidelity we calculate $G\left(n\right)$ in
continuous time for a harmonic well.

\subsection{Correlation of Wigner functions for different times in absence of
interactions}

Let $\omega_{1}$ be the frequency of a harmonic oscillator. The Wigner
function of a coherent state of the oscillator (\ref{eq: Coherent state wave function}),
corresponding to a time $t$ is (\ref{eq:Wigner function, harm. osc. general form}),
namely

\begin{equation}
W_{t}\left(x,p\right)=\frac{1}{2\pi\sigma_{x}\sigma_{p}}e^{-\frac{1}{2}\left(\frac{\left(x-x\left(t\right)\right)^{2}}{\sigma_{x}^{2}}+\frac{\left(p-p\left(t\right)\right)}{\sigma_{p}^{2}}\right)}.\label{eq:Wigner function of the oscillator, t1}
\end{equation}

\noindent The Wigner function corresponding to a time $t-\Delta t$
is

\noindent 
\begin{equation}
W_{t-\Delta t}\left(x,p\right)=\frac{1}{2\pi\sigma_{x}\sigma_{p}}e^{-\frac{1}{2}\left(\frac{\left(x-x\left(t-\Delta t\right)\right)^{2}}{\sigma_{x}^{2}}+\frac{\left(p-p\left(t-\Delta t\right)\right)^{2}}{\sigma_{p}^{2}}\right)},\label{eq:Wigner function of the oscillatior, t2}
\end{equation}
where $\sigma_{x}$ and $\sigma_{p}$ are given by (\ref{eq:width of Wigner function, x})
and (\ref{eq:width of Wigner function, p}) and denote the width of
the Wigner function in position and momentum, respectively. The difference
between the times is constant and given by $\Delta t.$

\noindent The correlation in absence of interactions is of the form

\begin{equation}
G=C\cdot e^{-\frac{1}{2}\left(s_{x}+s_{p}\right)}\label{eq:Correlation without interactions, general form}
\end{equation}

\noindent with

\begin{equation}
C=\frac{1}{4\pi\sigma_{x}\sigma_{p}},\label{eq:C}
\end{equation}

\begin{equation}
s_{x}=\frac{\left(x\left(t\right)-x\left(t-\Delta t\right)\right)^{2}}{2\sigma_{x}^{2}}
\end{equation}

\noindent and

\begin{equation}
s\left(p\right)=\frac{\left(p\left(t\right)-p\left(t-\Delta t\right)\right)^{2}}{2\sigma_{p}^{2}}.
\end{equation}

\noindent The phase coordinates are

\begin{equation}
\left(x\left(t\right),p\left(t\right)\right)=\rho\left(\cos\left(\omega t\right),-m\omega\sin\left(\omega t\right)\right)
\end{equation}
and

\begin{equation}
\left(x\left(t-\Delta t\right),p\left(t-\Delta t\right)\right)=\rho\left(\cos\left(\omega t+\phi\right),-m\omega\sin\left(\omega t+\phi\right)\right),
\end{equation}

where

\begin{equation}
\phi=\omega\Delta t.
\end{equation}

\subsection{Correlation of Wigner functions for different times with weak interactions}

The width of the Wigner functions in presence of weak interactions
is given by

\begin{equation}
\tilde{\sigma}_{x_{1}}^{2}=\sigma_{x}^{2}+\gamma_{x}\cos\left(\Omega t+\phi_{x}\right),
\end{equation}

\begin{equation}
\tilde{\sigma}_{x_{2}}=\sigma_{x}^{2}+\gamma_{x}\cos\left(\Omega t+\phi_{x}-\Delta\phi\right),
\end{equation}

\begin{equation}
\tilde{\sigma}_{p_{1}}=\sigma_{p}^{2}+\gamma_{p}\cos\left(\Omega t+\phi_{p}\right)
\end{equation}

and

\begin{equation}
\tilde{\sigma}_{p_{2}}^{2}=\sigma_{p}^{2}+\gamma_{p}\cos\left(\Omega t+\phi_{p}-\Delta\phi\right),
\end{equation}

\noindent where 

\begin{equation}
\Delta\phi=\Omega\cdot\Delta t.\label{eq:delta_phi}
\end{equation}

Therefore, 

\begin{equation}
C\left(t\right)=\left(2\pi\sigma_{x\left(t\right)}\sigma_{p\left(t\right)}\sigma_{x\left(t-\Delta t\right)}\sigma_{p\left(t-\Delta t\right)}\right)^{-1}\sqrt{\frac{\sigma_{x\left(t\right)}^{2}\sigma_{x\left(t-\Delta t\right)}^{2}\sigma_{p\left(t\right)}^{2}\sigma_{p\left(t-\Delta t\right)}^{2}}{\left(\sigma_{x\left(t\right)}^{2}+\sigma_{x\left(t-\Delta t\right)}^{2}\right)\left(\sigma_{p\left(t\right)}^{2}+\sigma_{p\left(t-\Delta t\right)}^{2}\right)}}\label{eq:C_interactions}
\end{equation}

and

\begin{equation}
G\left(t\right)=C\left(t\right)\cdot e^{-\frac{1}{2}\left(s_{x}\left(t\right)+s_{p}\left(t\right)\right)}.\label{eq:Correlation without interactions, general form-1}
\end{equation}

The expressions for $s\left(x\right)$ and $s\left(p\right)$ become

\begin{equation}
s_{x}\left(t\right)=\frac{\rho^{2}}{2\sigma_{x}^{2}}\cdot\frac{\left(\cos\left(\omega t\right)-\cos\left(\omega t-\phi\right)\right)^{2}}{1+\gamma_{x}\left(\cos\left(\Omega t+\phi_{x}\right)+\cos\left(\Omega t+\phi_{x}-\Delta\phi\right)\right)}\label{eq:s(t1)}
\end{equation}

and

\begin{equation}
s_{p}\left(t\right)=\frac{\rho^{2}m^{2}\omega^{2}}{2\sigma_{p}^{2}}\cdot\frac{\left(\sin\left(\omega t\right)-\sin\left(\omega t-\phi\right)\right)^{2}}{1+\gamma_{x}\left(\cos\left(\Omega t+\phi_{p}\right)+\cos\left(\Omega t+\phi_{p}-\Delta\phi\right)\right)}.
\end{equation}

Using (\ref{eq:cosine trig formula}), we get

\begin{equation}
s_{x}\left(t\right)=\sum_{i=1}^{8}A_{i},
\end{equation}

\noindent where

\begin{equation}
A_{1}=\frac{\rho^{2}}{2\sigma_{x}^{2}}-\frac{\rho^{2}\gamma_{x}}{2\sigma_{x}^{2}}\cos\left(\Omega t+\phi_{x}\right)-\frac{\rho^{2}\gamma_{x}}{2\sigma_{x}^{2}}\cos\left(\Omega t+\phi_{x}-\Delta\phi\right),
\end{equation}

\begin{equation}
A_{2}=\frac{\rho^{2}}{4\sigma_{x}^{2}}\cos\left(2\omega t\right)+\frac{\rho^{2}}{4\sigma_{x}^{2}}\cos\left(2\omega t-2\phi\right)-\frac{\rho^{2}}{2\sigma_{x}^{2}}\cos\left(\phi\right)\cos\left(2\omega t-\phi\right),
\end{equation}

\begin{equation}
A_{3}=-\frac{\rho^{2}\gamma_{x}}{8\sigma_{x}^{2}}\cos\left(\left(2\omega-\Omega\right)t-\phi_{x}\right)-\frac{\rho^{2}\gamma_{x}}{8\sigma_{x}^{2}}\cos\left(\left(2\omega-\Omega\right)t-\phi_{x}+\Delta\phi\right),\label{eq:A3_1}
\end{equation}

\begin{equation}
A_{4}=-\frac{\rho^{2}\gamma_{x}}{8\sigma_{x}^{2}}\cos\left(\left(2\omega-\Omega\right)t-2\phi-\phi_{x}+\Delta\phi\right)+\frac{\rho^{2}\gamma_{x}}{4\sigma_{x}^{2}}\cos\left(\left(2\omega-\Omega\right)t-\phi-\phi_{x}\right),\label{eq:A4_1}
\end{equation}

\begin{equation}
A_{5}=\frac{\rho^{2}\gamma_{x}}{4\sigma_{x}^{2}}\cos\left(\phi\right)\cos\left(\left(2\omega-\Omega\right)t-\phi-\phi_{x}+\Delta\phi\right)-\frac{\rho^{2}\gamma_{x}}{8\sigma_{x}^{2}}\cos\left(\left(2\omega-\Omega\right)t-2\phi-\phi_{x}\right),\label{eq:A5_1}
\end{equation}

\begin{equation}
A_{6}=-\frac{\rho^{2}\gamma_{x}}{8\sigma_{x}^{2}}\cos\left(\left(2\omega+\Omega\right)t+\phi_{x}\right)-\frac{\rho^{2}\gamma_{x}}{8\sigma_{x}^{2}}\cos\left(\left(2\omega+\Omega\right)t+\phi_{x}-\Delta\phi\right),
\end{equation}

\begin{equation}
A_{7}=-\frac{\rho^{2}\gamma_{x}}{8\sigma_{x}^{2}}\cos\left(\left(2\omega+\Omega\right)t-2\phi+\phi_{x}\right)-\frac{\rho^{2}\gamma_{x}}{8\sigma_{x}^{2}}\cos\left(\left(2\omega+\Omega\right)t-2\phi+\phi_{x}-\Delta\phi\right)
\end{equation}
and

\begin{equation}
A_{8}=\frac{\rho^{2}\gamma_{x}}{4\sigma_{x}^{2}}\cos\left(\phi\right)\cos\left(\left(2\omega+\Omega\right)t-\phi+\phi_{x}\right)+\frac{\rho^{2}\gamma_{x}}{4\sigma_{x}^{2}}\cos\left(\phi\right)\cos\left(\left(2\omega+\Omega\right)t-\phi+\phi_{x}-\Delta\phi\right).
\end{equation}

The intermediate frequency is present in (\ref{eq:A3_1}) - (\ref{eq:A5_1})
and is equal to $\Delta\omega=2\omega-\Omega.$ 

Similarly, for $s_{p}\left(t\right)$, 

\begin{equation}
s_{p}\left(t\right)=\sum_{i=1}^{8}A_{i},
\end{equation}

where

\begin{equation}
A_{1}=\frac{\rho^{2}m^{2}\omega^{2}}{2\sigma_{p}^{2}}\left(1-\cos\left(\phi\right)\right)-\frac{\rho^{2}m^{2}\omega^{2}\gamma_{p}}{2\sigma_{p}^{2}}\cos\left(\Omega t+\phi_{p}\right)-\frac{\rho^{2}m^{2}\omega^{2}\gamma_{p}}{2\sigma_{p}^{2}}\cos\left(\Omega t+\phi_{p}-\Delta\phi\right),
\end{equation}

\begin{equation}
A_{2}=\frac{\rho^{2}m^{2}\omega^{2}\gamma_{p}}{2\sigma_{p}^{2}}\cos\left(\phi\right)\cos\left(\Omega t+\phi_{p}\right)+\frac{\rho^{2}m^{2}\omega^{2}\gamma_{p}}{2\sigma_{p}^{2}}\cos\left(\phi\right)\cos\left(\Omega t+\phi_{p}-\Delta\phi\right),
\end{equation}

\begin{equation}
A_{3}=-\frac{\rho^{2}m^{2}\omega^{2}}{2\sigma_{p}^{2}}\cos\left(2\omega t\right)-\frac{\rho^{2}m^{2}\omega^{2}}{2\sigma_{p}^{2}}\cos\left(2\omega t+2\phi\right)+\frac{\rho^{2}m^{2}\omega^{2}}{2\sigma_{p}^{2}}\cos\left(2\omega t+\phi\right),
\end{equation}

\begin{equation}
A_{4}=\frac{\rho^{2}m^{2}\omega^{2}\gamma_{p}}{4\sigma_{p}^{2}}\cos\left(\left(2\omega-\Omega\right)t-\phi_{p}\right)+\frac{\rho^{2}m^{2}\omega^{2}\gamma_{p}}{4\sigma_{p}^{2}}\cos\left(\left(2\omega-\Omega\right)t-\phi_{p}+\Delta\phi\right),\label{eq:A4_2}
\end{equation}

\begin{equation}
A_{5}=\frac{\rho^{2}m^{2}\omega^{2}\gamma_{p}}{4\sigma_{p}^{2}}\cos\left(\left(2\omega-\Omega\right)t+2\phi-\phi_{p}\right)+\frac{\rho^{2}m^{2}\omega^{2}\gamma_{p}}{4\sigma_{p}^{2}}\cos\left(\left(2\omega-\Omega\right)t+2\phi-\phi_{p}+\Delta\phi\right),\label{eq:A5_2}
\end{equation}

\begin{equation}
A_{6}=-\frac{\rho^{2}m^{2}\omega^{2}\gamma_{p}}{4\sigma_{p}^{2}}\cos\left(\left(2\omega-\Omega\right)t+\phi-\phi_{p}\right)-\frac{\rho^{2}m^{2}\omega^{2}\gamma_{p}}{4\sigma_{p}^{2}}\cos\left(\left(2\omega-\Omega\right)t+\phi-\phi_{p}\Delta\phi\right),\label{eq:A6_2}
\end{equation}

\begin{equation}
A_{7}=\frac{\rho^{2}m^{2}\omega^{2}\gamma_{p}}{4\sigma_{p}^{2}}\cos\left(\left(2\omega+\Omega\right)t+\phi_{p}\right)+\frac{\rho^{2}m^{2}\omega^{2}\gamma_{p}}{4\sigma_{p}^{2}}\cos\left(\left(2\omega+\Omega\right)t+\phi_{p}-\Delta\phi\right),
\end{equation}

\begin{equation}
A_{8}=\frac{\rho^{2}m^{2}\omega^{2}\gamma_{p}}{4\sigma_{p}^{2}}\cos\left(\left(2\omega+\Omega\right)t+2\phi+\phi_{p}\right)+\frac{\rho^{2}m^{2}\omega^{2}\gamma_{p}}{4\sigma_{p}^{2}}\cos\left(\left(2\omega+\Omega\right)t+2\phi+\phi_{p}-\Delta\phi\right)
\end{equation}

and

\begin{equation}
A_{9}=-\frac{\rho^{2}m^{2}\omega^{2}\gamma_{p}}{4\sigma_{p}^{2}}\cos\left(\left(2\omega+\Omega\right)t+\phi+\phi_{p}\right)-\frac{\rho^{2}m^{2}\omega^{2}\gamma_{p}}{4\sigma_{p}^{2}}\cos\left(\left(2\omega+\Omega\right)t+\phi+\phi_{p}-\Delta\phi\right).
\end{equation}

The intermediate frequency $\Delta\omega=2\omega-\Omega$ can be seen
in (\ref{eq:A4_2}) - (\ref{eq:A6_2}). 

\bibliographystyle{IEEEtran}
\bibliography{Reference_database1}

\begin{thebibliography}{10}
\providecommand{\url}[1]{#1}
\csname url@samestyle\endcsname
\providecommand{\newblock}{\relax}
\providecommand{\bibinfo}[2]{#2}
\providecommand{\BIBentrySTDinterwordspacing}{\spaceskip=0pt\relax}
\providecommand{\BIBentryALTinterwordstretchfactor}{4}
\providecommand{\BIBentryALTinterwordspacing}{\spaceskip=\fontdimen2\font plus
\BIBentryALTinterwordstretchfactor\fontdimen3\font minus
  \fontdimen4\font\relax}
\providecommand{\BIBforeignlanguage}[2]{{%
\expandafter\ifx\csname l@#1\endcsname\relax
\typeout{** WARNING: IEEEtran.bst: No hyphenation pattern has been}%
\typeout{** loaded for the language `#1'. Using the pattern for}%
\typeout{** the default language instead.}%
\else
\language=\csname l@#1\endcsname
\fi
#2}}
\providecommand{\BIBdecl}{\relax}
\BIBdecl

\bibitem{Quantum_accelerator}
L.Rebuzzini, R.Artuso, S.Fishman, and I.~Guarneri, ``Effects of atomic
  interactions on quantum accelerator modes,'' \emph{Phys. Rev. A}, vol. 76,
  031603(R), 2007.

\bibitem{Transition_to_instability}
C.~Zhang, J.~Liu, M.~G.Raizen, and Q.~Niu, ``Transition to instability in a
  kicked bose-einstein condensate,'' \emph{Phys. Rev. Lett.}, vol.~92, p.
  054101, 2004.

\bibitem{Wave_packet_dynamics}
S.Moulieras, A.G.Monastra, M.Saraceno, and P.Leboeuf, ``Wave-packet dynamics in
  nonlinear schrodinger equations,'' \emph{Phys. Rev. A}, vol. 85, 013841,
  2012.

\bibitem{Mark_paper}
M.Herrera, T.M.Antonsen, E.Ott, and S.Fishman, ``Dynamic localization of a
  weakly interacting bose-einstein condensate in an anharmonic potential,''
  \emph{Phys. Rev. A}, vol. 87,041603(R), 2013.

\bibitem{Peres_fidelity}
A.Peres, ``Stability of quantum motion in chaotic and regular systems,''
  \emph{Phys. Rev. A}, vol. 30, 1610, 1984.

\bibitem{Jalabert_Pastawski_Y19}
R.A.Jalabert and H.M.Pastawski, ``Environmental-independent decoherence rate in
  classically chaotic systems,'' \emph{Phys. Rev. Lett.}, vol. 86, 2490, 2001.

\bibitem{Decay_Loschmidt_Y20}
P.Jacquod, I.Adagideli, and C.W.J.Beenakker, ``Decay of the loschmidt echo for
  quantum states with sub-planck-scale structures,'' \emph{Phys. Rev. Lett.},
  vol. 89, 154103, 2002.

\bibitem{Sensitivity_Chaotic_Systems_Y34}
N.R.Cerruti and S.Tomsovic, ``Sensitivity of wave field evolution and manifold
  stability in chaotic systems,'' \emph{Phys. Rev. Lett.}, vol. 88, 054103,
  2002.

\bibitem{Saturation_of_Fidelity_Y35}
S.Wimberger and A.~Buchleitner, ``Saturation of fidelity in the atom-optics
  kicked rotor,'' \emph{J. Phys. B}, vol. 39, L145, 2006.

\bibitem{Decay_of_QM_correlations_Y38}
M.F.Andersen, A.Kaplan, T.~Grunzweig, and N.Davidson, ``Decay of quantum
  correlations in atom optic billiards with chaotic and mixed dynamics,''
  \emph{Phys. Rev. Lett.}, vol. 97, 104102, 2006.

\bibitem{Hyperfine_spectroscopy_Y37}
A.Kaplan, M.Andersen, T.Grunzweig, and N.Davidson, ``Hyperfine spectroscopy of
  optically trapped atoms,'' \emph{J. Opt. B}, vol. 7, R103, 2005.

\bibitem{Revivals_of_coherence_Y36}
M.F.Andersen, T.Grunzweig, A.Kaplan, and N.Davidson, ``Revivals of coherence in
  chaotic atom-optics billiards,'' \emph{Phys. Rev. A}, vol. 69, 063413, 2004.

\bibitem{Review_Y21}
T.Gorin, T.Prosen, T.H.Seligman, and M.Znidaric, ``Dynamics of loschmidt echoes
  and fidelity decay,'' \emph{Phys. Rep.}, vol. 435, 33, 2006.

\bibitem{Yevgeny_paper}
Y.~Krivolapov, S.~Fishman, E.~Ott, and T.~M. Antonsen, ``Quantum chaos of a
  mixed open system of kicked cold atoms,'' \emph{Phys. Rev. E}, vol.
  83,016204, 2011.

\bibitem{Mark_Update}
M.Herrera, ``private communication,'' april 2013.

\bibitem{Shepelyansky_93}
D.L.Shepelyansky, ``Delocalization of quantum chaos by weak nonlinearity,''
  \emph{Phys. Rev. Lett.}, vol. 70, 1787, 1993.

\bibitem{Echoes}
M.Herrera, T.M.Antonsen, E.Ott, and S.Fishman, ``Echoes and revival echoes in
  systems of anharmonically confined atoms,'' \emph{Phys. Rev. A}, vol.
  86,023613, 2012.

\bibitem{coherent_comp}
P.Milloni and M.Nieto, ``Coherent states,'' in \emph{Compendium of Quantum
  Physics}, D.~Greenberger, K.~Hentschel, and F.~Weinert, Eds.\hskip 1em plus
  0.5em minus 0.4em\relax Springer Berlin Heidelberg, 2009, pp. 106--108.

\end{thebibliography}

\end{document}